\documentclass[printer]{aa} 
\usepackage{graphicx}
\usepackage{natbib}
\usepackage{amsmath}
\usepackage{amscd}
\usepackage{booktabs}
\usepackage{caption}
\usepackage{multirow}
\usepackage{subfigure}
\usepackage{url}
\usepackage{bm}
\usepackage{color}
\usepackage[usenames,dvipsnames,svgnames,table]{xcolor}
\usepackage{txfonts}
\usepackage{lineno}
\bibpunct{(}{)}{;}{a}{}{,} 

\begin{document}
   \title{Cluster membership probabilities from proper motions and
     multiwavelength photometric catalogues:}
   \subtitle{I. Method and application to the Pleiades cluster}
   \author{
     L. M. Sarro\inst{1,2}
     \and H. Bouy\inst{3}
     \and A. Berihuete\inst{4}
     \and E. Bertin\inst{5}
     \and E. Moraux\inst{6}
     \and J. Bouvier\inst{6}
     \and J.-C. Cuillandre\inst{7}
     \and D. Barrado\inst{3}
     \and E. Solano\inst{3,2}
   }

   \institute{
     Dpto. de Inteligencia Artificial, ETSI Inform\'atica, UNED, Juan del Rosal, 16,
     E-28040, Madrid, Spain\\
     \and
     Spanish Virtual Observatory\\
     \and
     Centro de Astrobiolog\'\i a, depto. de Astrof\'\i sica, 
     INTA-CSIC, PO BOX 78, E-28691, ESAC Campus, 
     Villanueva de la Ca\~nada, Madrid, Spain\\
     \and
     Dpt. Statistics and Operations Research, University of
     C\'adiz, Campus Universitario R\'io San Pedro s/n.  11510 Puerto
     Real, C\'adiz, Spain\\
     \and
     Institut d'Astrophysique de Paris, 
     CNRS UMR 7095 and UPMC, 98bis bd Arago, 
     F-75014 Paris, France \\
     \and
     UJF-Grenoble 1/CNRS-INSU, 
     Institut de Plan\'etologie et d'Astrophysique de Grenoble (IPAG), 
     UMR 5274, Grenoble, F-38041, France\\
     \and
     Canada-France-Hawaii Telescope Corporation, 
     65-1238 Mamalahoa Highway, Kamuela, HI96743, USA \\
   }    

   \date{}
 
  \abstract
  {With the advent of deep wide surveys, large photometric and
    astrometric catalogues of literally all nearby clusters and
    associations have been produced. The unprecedented accuracy and
    sensitivity of these data sets and their broad spatial, temporal
    and wavelength coverage make obsolete the classical membership
    selection methods that were based on a handful of colours and
    luminosities. We present a new technique designed to take full
    advantage of the high dimensionality (photometric, astrometric,
    temporal) of such a survey to derive self-consistent and robust
    membership probabilities of the Pleiades cluster.}
  {We aim at developing a methodology to infer membership
    probabilities to the Pleiades cluster from the DANCe
    multidimensional astro-photometric data set in a consistent way
    throughout the entire derivation. The determination of the membership
    probabilities has to be applicable to censored data and must
    incorporate the measurement uncertainties into the inference
    procedure.}
  {We use Bayes' theorem and a curvilinear forward model for the
    likelihood of the measurements of cluster members in the
    colour-magnitude space, to infer posterior membership
    probabilities. The distribution of the cluster members proper
    motions and the distribution of contaminants in the full
    multidimensional astro-photometric space is modelled with a
    mixture-of-Gaussians likelihood.}
 {We analyse several representation spaces composed of the proper
   motions plus a subset of the available magnitudes and colour
   indices. We select two prominent representation spaces composed of
   variables selected using feature relevance determination techniques
   based in Random Forests, and analyse the resulting samples of high
   probability candidates. We consistently find lists of high
   probability ($p >$ 0.9975) candidates with $\approx$ 1000 sources,
   4 to 5 times more than obtained in the most recent astro-photometric 
   studies of the cluster.}
  {Multidimensional data sets require statistically sound
    multivariate analysis techniques to fully exploit
    their scientific information content. Proper motions in particular
    are, as expected, critical for the correct separation of
    contaminants. The methodology presented here is ready for
    application in data sets that include more dimensions, such as
    radial and/or rotational velocities, spectral indices and
    variability.}
   \maketitle
%


\section{Introduction}
\label{intro}

The analysis of stellar clusters is one of the stepping stones for
understanding galactic and stellar formation and evolution. The study
of different clusters reveals properties of the probabilistic
distribution of initial masses (the initial mass function, IMF) and
its dependence on parameters such as metallicity and age. The
evolution of their dynamical scales and properties can help us
understand the history of our own Galaxy and its components. But using
clusters as one of the building blocks of our knowledge of the nearer
Universe requires identifications of the cluster members as completely
and accurately as possible. This work is the first in a series where we
apply recent developments in statistics to the problem of estimating
the cluster membership probabilities for a set of sources. We
concentrate on data sets with both proper motions and multiband
photometry.

Traditionally, the problem of estimating membership probabilities
based on proper motion measurements has been treated with techniques
that followed the pioneering work by \cite{1958AJ.....63..387V} and
\cite{1971A&A....14..226S}. These works modelled the distribution of
sources in the vector point diagram (hereafter VPD) as the mixture of
two bi-variate Gaussian distributions, one for the cluster members and
another for the field sources. The cluster distribution was assumed to be
circular in both cases because the astrometric uncertainties did not
allow for the detailed resolution of the real shape, and in both cases
the membership probability is defined as the posterior probability of
the class given the data in this simplified model. Both works adopt
a Bayesian approach to inferring membership probabilities, only in the
sense that they apply Bayes' theorem. With Bayes' theorem, the
posterior probability of the unknown parameter ${\cal C}$ (the class
of the source, i.e., member of the cluster or field populations) given
the observed data ${\cal D}$ is expressed in terms of a model of the
probabilistic distribution of sources of each class in the space of
the observed data ${\cal D}$, an {\sl a priori} membership probability
to each class, and a normalising constant:

\begin{equation}
p({\cal C}|{\cal D}) = \frac{p({\cal D}|{\cal C})\cdot p({\cal
    C})}{p({\cal D})}.
\label{bayes.th}
\end{equation}

We use ${\cal C}$, a dichotomic random variable, to represent
membership to the foreground and background populations (${\cal
  C}_1$) or to the stellar cluster (${\cal C}_2$). In the following,
the subscript $i$ indexes these two populations, and we use ${\cal
  C}_i$ as an abbreviation of ${\cal C}={\cal C}_i$.

In Eq. \ref{bayes.th}, the posterior membership probability (the
left-hand side) is expressed in terms of the likelihood ($p(D|{\cal
  C})$) and the prior probabilities ($p({\cal C})$ and $p({\cal
  D})$). It is precisely the likelihood that is modelled in these
pioneering works as a symmetric Gaussian distribution in ${\cal D}$,
the space of proper motions.

\cite{1995AJ....109..672K} combined celestial coordinates with proper
motions assuming conditional independency to refine their
membership probabilities. They followed the conceptual framework set up
by the earlier works cited in the previous paragraph, based on the
posterior probability density distribution of a bi-variate Gaussian
mixture model. Additionally and alternatively, they proposed an
improvement of an original procedure introduced in
\cite{1991A&AS...87...69P}, where the modelling procedure was treated
by partitioning the data into moving bins in a data space that
comprises celestial coordinates, proper motions, and the B
magnitude. Within each bin, the field distribution is assumed to be a
flat linear function in the space of proper motions. The membership
probability is then derived using Bayes' theorem locally for each
target star in each bin, fixing the mean of the Gaussian distribution
in the vector point diagram (VPD) and fitting the prior probabilities
with the least-squares estimate in a 8$\sigma\times8\sigma$ square
(with $\sigma$ being the fixed standard deviation of the Gaussian
distribution). We recall here that the least-squares fit is equivalent
to the maximum-likelihood solution under the assumption of independent
Gaussian uncertainties with constant standard deviation. Hence the
grouping of the data in magnitude bins. This approach (binning in
magnitude) became popular in later works \citep[see][for a recent
  example]{2012MNRAS.422.1495L}. This latter alternative in the work
by \cite{1995AJ....109..672K} presents several disadvantages in our
opinion, including the occasional introduction of faint stars or
fractions thereof ({\sl sic}) to avoid negative values of the
field-star distribution.

An alternative to this parametric modelling of the distribution of
cluster and field members in the VPD was proposed by
\cite{2007A&A...470..585B}, who derived kernel-based
estimations of the probability density in the VPD space both for field
stars alone (in a region assumed free from cluster members) and for
the mixed population in the cluster celestial region. 

Later works with combined astrometric and spectro-photometric data
sets elaborated further on these early works. Broad-band photometry is
usually considered separately from the astrometric measurements,
either before or after the astrometric analysis, and as hard selection
thresholds in colour-magnitude or colour-colour diagrams (hereafter
CMD and CCD respectively; see for example \cite{2004A&A...416..125D}
or \cite{2007A&A...470..585B}). Usually, these thresholds do not
incorporate the photometric uncertainties in a way that retains the
probabilistic nature of the memberships, and consist of the refinement
of the membership list by removing candidates outside certain bands
around the cluster main sequence as defined by theoretical
evolutionary tracks. These cuts are also applied in other
representation sub-spaces such as derived from radial velocity,
spectral type, gravity, activity, metallicity, or lithium line
measurements.

The aforecited work by \cite{2012MNRAS.422.1495L} and the companion
articles \cite{2013MNRAS.431.3222L}, \cite{2012MNRAS.426.3419B}, and
\cite{2012MNRAS.426.3403L}, are recent examples of the classical
methodology to derive membership lists from astro- and photometric
data sets, this time in the same stellar cluster as studied in the
current work: the Pleiades. \cite{2012MNRAS.422.1495L} defined the
cluster sequence in the (Z-J)--Z CMD from previously known members;
this sequence induces a conservative photometric cut in the original
data set that is meant to remove a large part of field sources;
finally, an initial list of membership probabilities is derived
following an incorrect version\footnote{There is a missing factor
  (1-f) in their Eq. 5. This factor should be multiplying the
  nominator, and is needed for $p$ to be bounded between 0 and 1, as
  formally required of a probability.} of Eq. \ref{bayes.th} applied
to the VPD representation space. A threshold of $p=0.6$ results in 947
member candidates.

The methodologies presented thus far do not extend the probabilistic
treatment first put forward by \cite{1958AJ.....63..387V} and
\cite{1971A&A....14..226S} for the VPD, to extended data sets with
photometric or spectrometric measurements. An exception to this is
the recent work by \cite{2013ApJ...762...88M} in young stellar
kinematic groups, where the authors treat the astrometric and
photometric data consistently in a probabilistic framework, albeit
assuming that all measurements included in the inference of membership
probabilities are independent and uncorrelated. In general, we find
that the application of magnitude-dependent hard thresholds in
consecutive CMDs has two main drawbacks: it renders the resulting
membership probabilities inconsistent across magnitude bins, and
produces membership probabilities only for sources with measurements
in each and every CMD used for selection.

\cite{2014A&A...561A..57K} have recently developed an unsupervised
method for computing membership probabilities from celestial
coordinates and photometric measurements using principal components
and clustering methods. Their methodology is related to the one
presented in this work, and is as well capable of incorporating models
of uncertainty and incomplete observations. As the authors point out,
one of the main disadvantages of their method is the linear projection
implicit to the principal components analysis. The distribution of
sources in the colour-magnitude space is non-linear in general, and a
linear projection inevitably results in membership probabilities that
are systematically biased \citep[see][for a proposal to circumvent
  this problem]{2014A&A...561A..57K}.

A final remark on the classical methodology for the determination of
membership probabilities in stellar clusters or groups is that the
aforecited works do not incorporate measurement uncertainties into the
derivation of the membership probabilities in a consistent way
throughout the entire magnitude and proper motion range. We aim at
producing membership probabilities with a method that treats
astrometric and photometric measurements and their uncertainties
consistently. We extend the original probabilistic formulation by
\cite{1958AJ.....63..387V} and \cite{1971A&A....14..226S} to include
magnitudes and colour indices as part of the representation space
where the modelling takes place, and incorporate all the measurement
uncertainties (i.e. both astro- and photometric) into the derivation
of probabilistic memberships. Another objective of this work is to
provide an inference system capable of handling incomplete data (that
is, data with non-detections, upper limits, or corrupt values).

In Section \ref{sect:rml}, we describe in more detail the general
framework of our methods and the data set that drives the necessity to
develop a methodology to infer membership probabilites. In Section
\ref{sect:repspace} we discuss the set of random variables used to
represent sources in the modelling of the distribution corresponding
to the cluster and the field, and alternative spaces tested in the
experimentation phase. In Section \ref{sect:results} we describe the
results obtained with two sets of models, and in Section
\ref{sect:sensitivity} we analyse the sensitivity of these results to
the various choices of parameters and representation space. Finally,
in Section \ref{sect:lim-concl} we expose the weaknesses and
limitations of this methodology and point out future developments that
circumvent these limitations.

\section{Membership probability estimation}
\label{sect:rml}

\subsection{Observations}

We explore probabilistic models for the estimation of membership
probabilities in a multidimensional data set composed of proper
motions and apparent magnitudes of sources in the Pleiades cluster sky
region. The origin of the data set and the typical uncertainties are
described in \cite{dance1}. The data set is composed of proper motions
in equatorial coordinates ($\mu_{\alpha}, \mu_{\delta}$) and
measurements in the u, g, r, i, Z, Y, J, H, and K bands for 3577478
sources. From these, we selected 2066111 sources with proper motions
between $\pm$ 100 mas yr$^{-1}$ (hereafter, the DANCe data
set). Unfortunately, the data set is inhomogeneous in the various
bands, with different spatial coverages, completenesses, and limiting
magnitudes \citep[see Fig. 1 in ][for a graphical representation of
  the complex spatial coverage of the DANCe Pleaides data
  set]{dance1}.

\begin{figure*}[thb]
  \centering
  \subfigure[]{\label{pm}
  \includegraphics[scale=0.12]{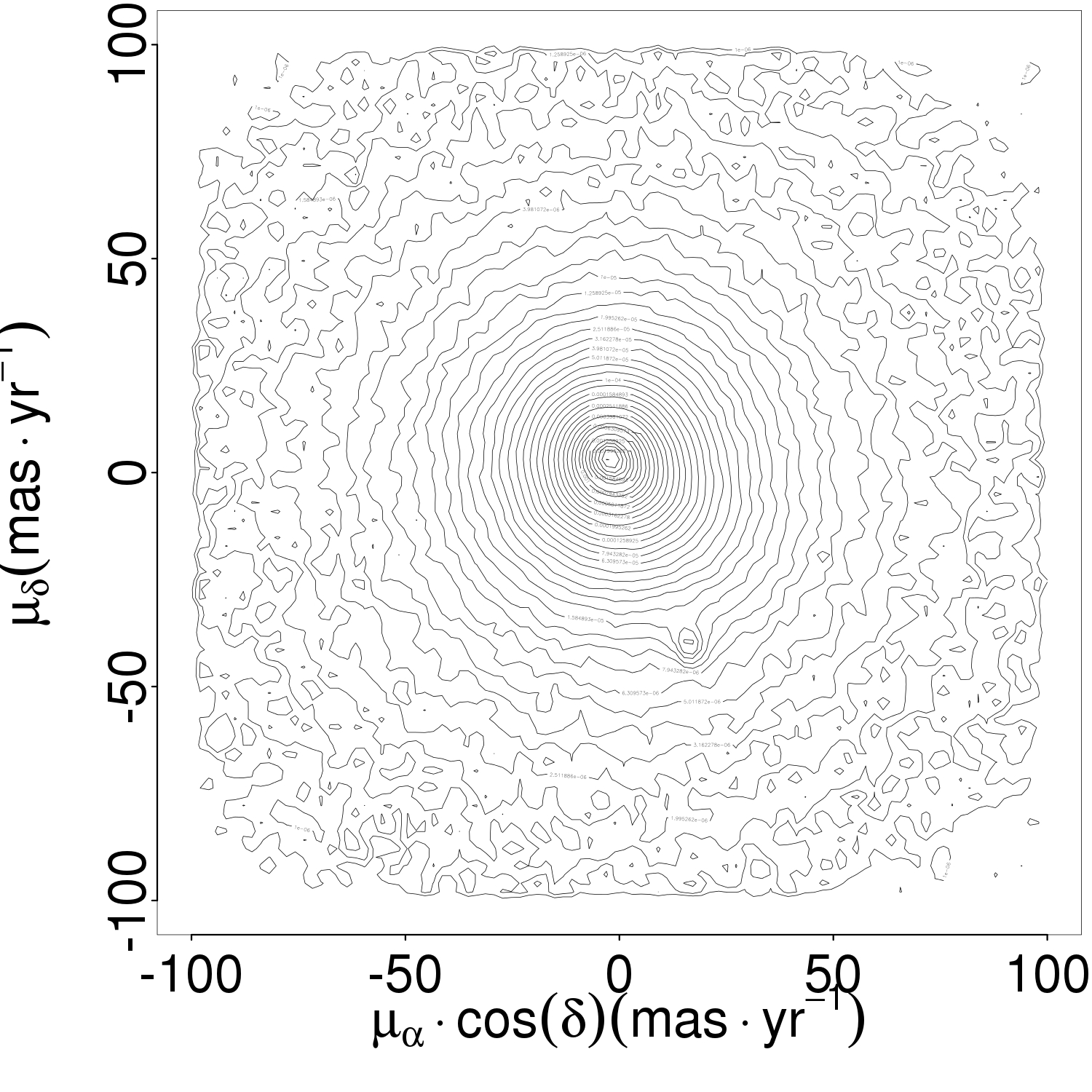}}
  \subfigure[]{\label{Kvsi-K}
  \includegraphics[scale=0.12]{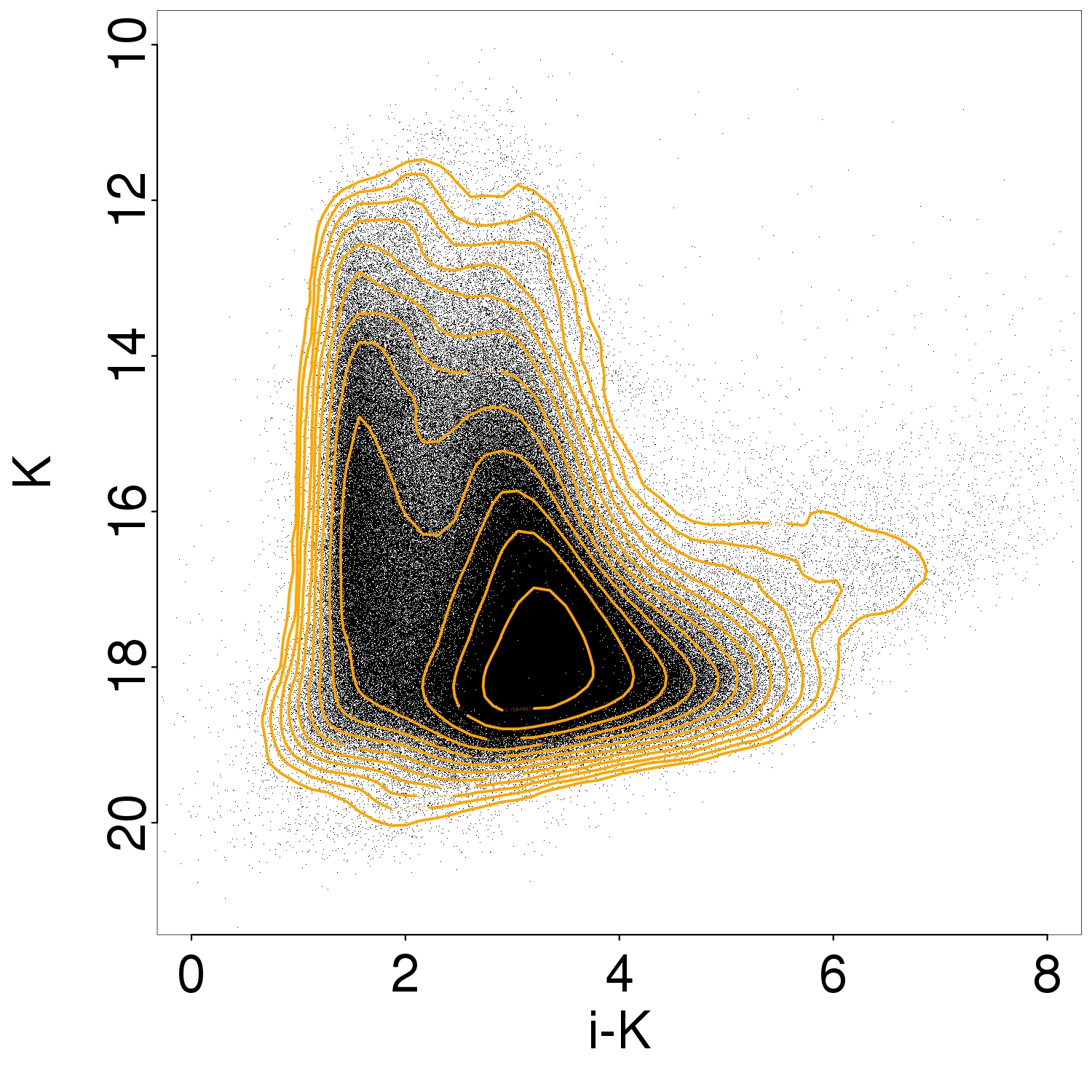}}
  \subfigure[]{\label{Kvsr-K}
  \includegraphics[scale=0.12]{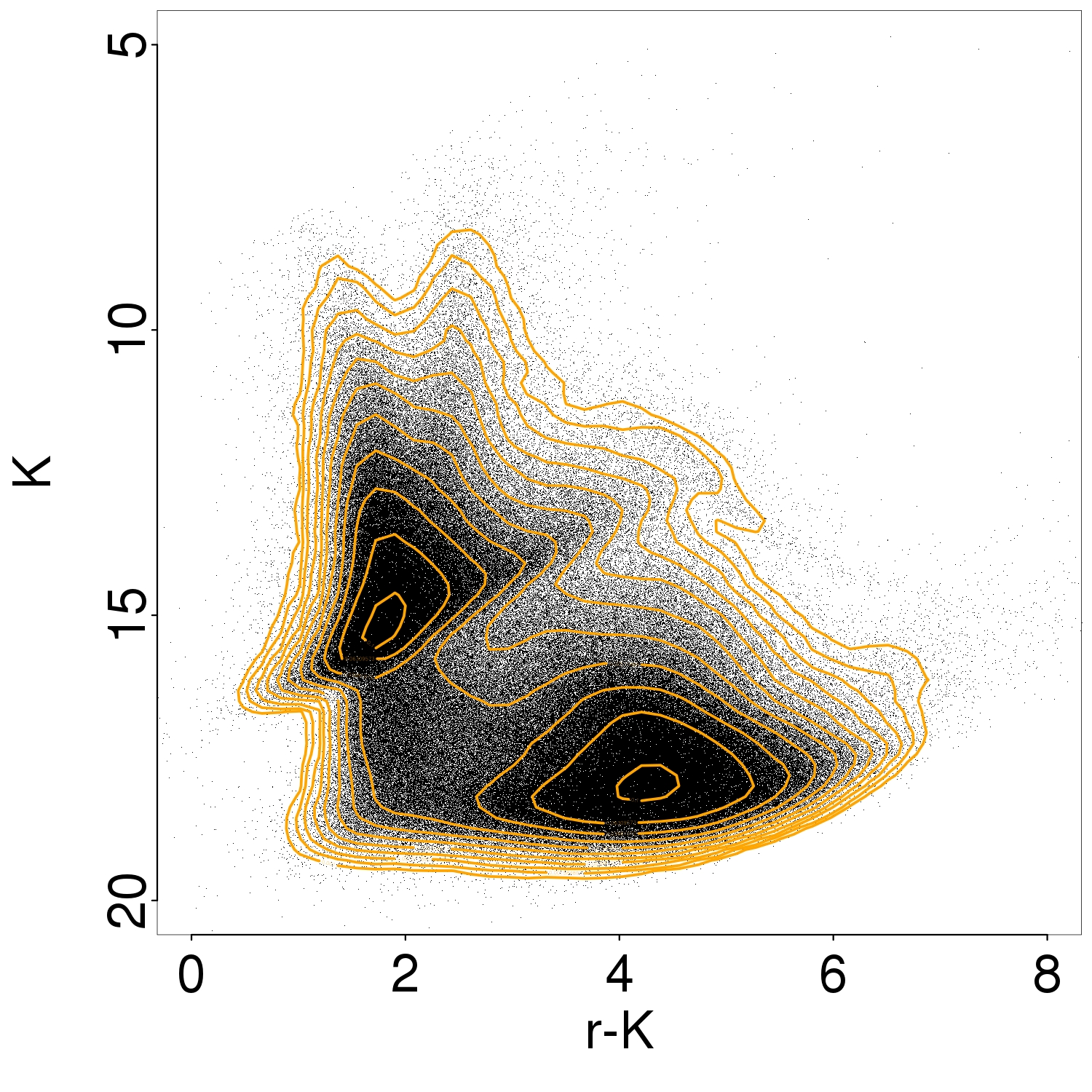}}
  \subfigure[]{\label{ivsi-K}
  \includegraphics[scale=0.12]{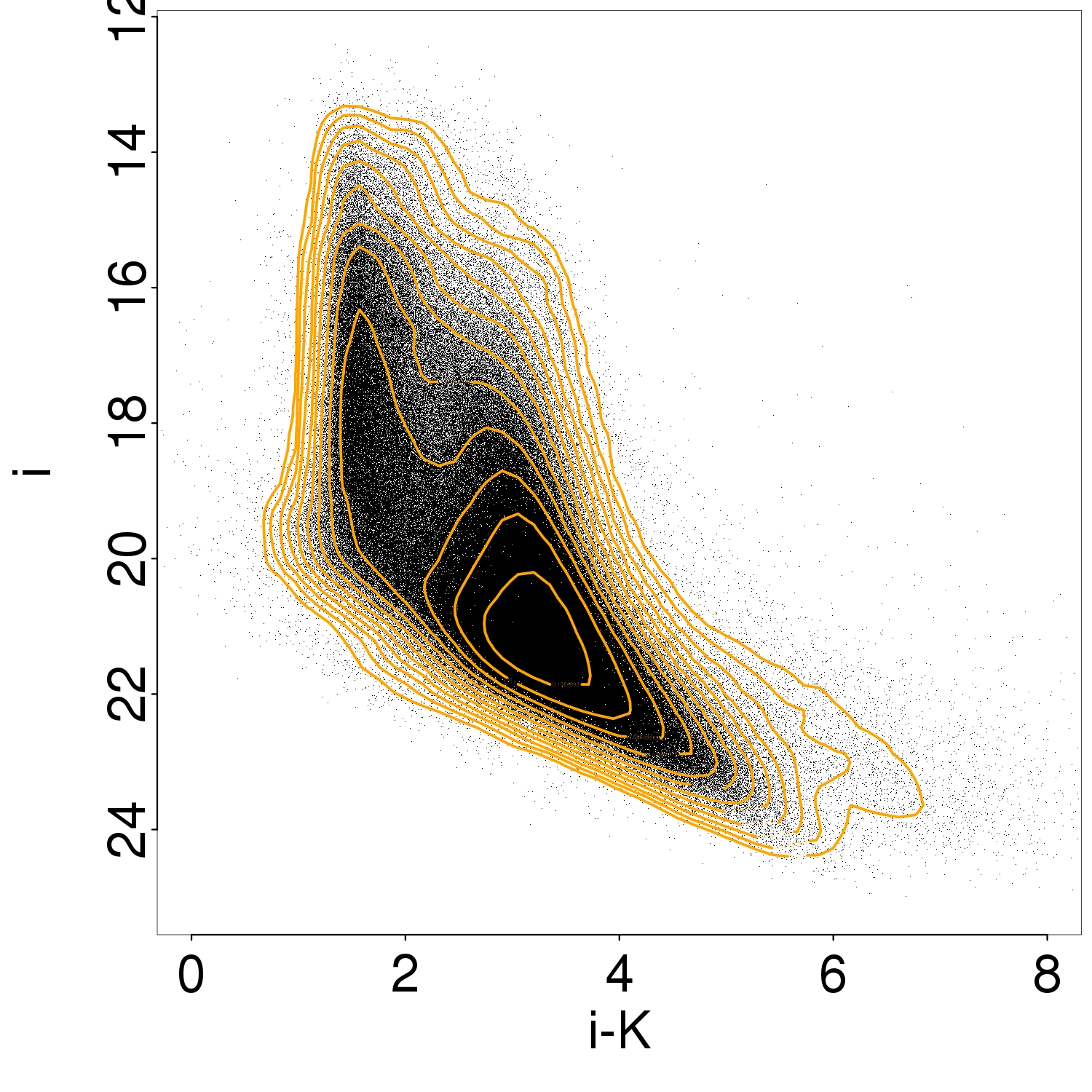}}
  \caption{\label{data} Original DANCe data set for the Pleiades
    cluster sky region in the VPD (a), (i-K)--K CMD (b), (r-K)--K CMD
    (c), and (i-K)--i CMD (d).}
\end{figure*}

As a consequence, only 132067 sources are complete (have photometric
measurements in all bands). Table \ref{nrofmeas} summarises the number
of missing measurements in each variable, and Fig. \ref{data} shows
the distribution of sources in the proper motion diagram or VPD (top
left) and several colour-magnitude sub-spaces (K versus
  (i-K), i versus (i-K), and K versus (r-K), from top right to bottom
  left clockwise).

\begin{table*}
\caption{Number of missing measurements}              
\label{nrofmeas}      
\centering                                      
\begin{tabular}{c  c  c  c  c  c  c  c  c  c  c}          
\hline                                             
$\mu_{\alpha}$ & $\mu_{\delta}$ & u & g & r & i & Z & Y & J & H & K \\    
\hline\hline                        
0  & 0 & 1842682 & 1627728 & 1294168  & 928902  & 725393  & 724041  & 601758  & 746009  & 768222\\

\hline                                             
\end{tabular}
\end{table*}

We used an initial list of Pleiades candidate members \citep[derived
  from][]{2007ApJS..172..663S} as the starting point
for the derivation of our models. We will refer to this initial list
as the initial reference set.

\subsection{Methodology}

The approach adopted here consists of inferring the posterior
membership probability given the observations $\mathcal{D}$, $p({\cal
  C}|{\cal D})$, using Bayes' theorem (Eq. \ref{bayes.th}).

In Eq. \ref{bayes.th}, the posterior membership probability (the
left-hand side) is expressed in terms of the likelihood ($p(D|{\cal
  C})$) and the prior probabilities ($p({\cal C})$ and $p({\cal
  D})$). We define the prior membership probability as
the proportion of cluster members in the entire data set.

In addition to the prior probabilities, a mathematical model of the
likelihood is needed to convert prior into posterior
probabilities. In the following paragraphs we explain our strategies
to derive the likelihood function. 

We modelled the distribution of examples of both members and
non-members in several representation spaces discussed below. For
example, a representation space can be composed of the two proper
motions, and a subset of magnitudes and colour indices. These are the
variables used to represent each source. In all cases, the likelihood
model was derived only from complete observations, that is, from
sources that have measurements available for all variables defining
the representation space. This likelihood model was then used in
Eq. \ref{bayes.th} to infer the membership probabilities of all
sources in the DANCe Pleiades data set.

We did not include in this likelihood model the spatial distribution
of cluster members in the celestial sphere. This is because one of the
main objectives of the analysis of the Pleaides DANCe dataset is to
infer this spatial distribution and study the statistical significance
of potential dependencies on the stellar mass. Thus, we prefered to
not assume any model for the cluster members spatial density, and
derived membership probabilities only from proper motions and
colour-magnitude diagrams. We postpone the inclusion of the spatial
information until a detailed analysis of the physical properties of
the cluster members is carried out and made public in a future article
in this series.

Both the initial list of Pleiades candidate members and the set of
DANCe sources in the field are the result of a spatial selection on
the celestial sphere and on the VPD diagram. Furthermore, since the
likelihood model is only infered from complete observations, the
spatial footprint of the different instruments included in the data
set represents yet another source of biases. The selection based on
celestial coordinates is defined by a circle of 5 degrees radius
around the centre of the cluster; the selection in the VPD is bounded
by maximum proper motions in any of the two axes of 100
mas$\cdot$yr$^{-1}$. This {\it ab initio} selection procedure defines
the proportion of the two types of sources (field and cluster) and
therefore affects the posterior probabilities via the likelihood and
prior probability, as indicated in Eq. \ref{bayes.th}. The posterior
probabilities derived in this work are thus not an absolute scale, but
are those derived from the actual data set, with the biases inherent
to the preparation of the DANCe catalogue.

In Sect. \ref{sect:lim-concl} we indicate a rigorous and consistent
way to take these selection biases into account; this involves
multilevel or hierarchical modelling of the data set.

\subsection{Mixture-of-Gaussians model for the field sources}
\label{MG}

We began by modelling the distribution of foreground and background 
sources in the representation space with a
mixture-of-Gaussians model. The mixture-of-Gaussians model is a simple
linear combination of $k_1$ multivariate normal (Gaussian)
distributions, with the index 1 refering to the set of field
contaminants (stars mostly, although it may contain one or
various types of extragalactic sources, an issue that will be pursued
further in future works).

A model is given by the $k_1$ means ($\vec{\mu}_{1,j}, j=1,2,...,k_1$)
and covariance matrices ($\Sigma_{1,j}, j=1,2,...,k_1$) of the
multivariate normal components in the mixture. We denote this set of
parameters as $\vec\theta_1$. Below we use the subscript $j$ for
indexing mixture components. Let $\vec{m}$ be the vector of
observations (the subset of proper motions, apparent magnitudes and
colour indices) used to represent sources. Then, the
mixture-of-Gaussians model of the likelihood for the field sources is
defined as

\begin{equation}
p(\vec{m}|{\cal C}_1)= \sum_{j=1}^{k_1} \pi_{1,j} \cdot {\cal
  N}_{\mathbf{\mu_{1,j}},\Sigma_{1,j}}(\vec{m}),
\end{equation}

with $\mathcal{N}$ representing the normal Gaussian probability
density, and $\pi_{1,j}$ the mixture proportions (i.e., the
probability that a given source belongs to the mixture component
$j$). 

We obtained maximum-likelihood models using the
expectation-maximisation (EM) algorithm for a sequence of $k_1$
values. The EM algorithm is not guaranteed to find the global maximum
of the likelihood landscape, but repeated runs with subsamples of the
complete data set did not result in large variations, either in the
number of components, or in the fitted parameters.

We denote the values of $\vec{\theta}_1^k$ that maximise the
likelihood function with $k$ Gaussian components as
$\vec{\hat\theta_1^k}$. ${\cal L}_{\hat\theta^k_1}$ is the maximum
likelihood found by the EM algorithm for the family of models defined
as a sum of $k$ multivariate normal distributions with full covariance
matrices, and the data set defined by $n$ data points (sources) of
dimension $d$. We then selected the likelihood model that maximises
the Bayesian information criterion (BIC), given by

\begin{equation}
{\cal M}= -2\ln({\cal L}_{\hat\theta_1^k})+k\frac{d(d+3)}{2}\ln(n).
\end{equation}

The BIC represents a regularised figure-of-merit with a penalty term
for complex models derived in the Bayesian framework for multivariate
distributions \citep{Schwarz1978Estimating}.

\subsection{Principal curve models for the cluster members}
\label{PC}

In Sect. \ref{MG} we modelled the likelihood term in
Eq. \ref{bayes.th} for the field sources as a sum of multivariate
normal distributions. To avoid the overfitting of the data, we used a
regularised maximum-likelihood criterion with a penalty for complex
models (i.e., the BIC). While this approach is a good starting point
for the complex distribution of field sources shown in
Fig. \ref{data}, for the cluster likelihood model we chose a different
class of models in which the probability density of the data in the
magnitude-colour subspace of the representation space can be modelled
by a curve. In this section, we construct a simplified model of the
distribution of cluster members in which the variables pertaining to
the proper motions and colour-magnitude subspaces are
independent. Therefore, the full likelihood model is decomposed as the
product of two models, one for each subspace. The model for the
subspace of proper motions still is a mixture-of-Gaussians, while the
likelihood model for the colour-magnitude (hereafter CM) subspace is a
curvilinear one based on principal curves, as described
below. Formally,

\begin{equation}
p(\vec{m}|{\cal C}_2)= \sum_{j=1}^{k_2} \pi_{2,j} \cdot {\cal
  N}_{\vec{\mu}_{2,j},\Sigma_{2,j}}(\mu_{\alpha},\mu_{\delta}) \cdot
p(\vec{m_{CM}}|{\cal C}_2),
\end{equation}

where $p(\vec{m}_{CM}|{\cal C}_2)$ is the curvilinear likelihood model
for the probability density distribution in the CM subspace.

The choice of an empirical curve over a theoretical isochrone frees us
from any assumptions on the cluster's parameters (e.g. age, distance,
metallicity) or from the model uncertainties, which are especially
important at young ages. We have also tested a single
mixture-of-Gaussians model for the sample of sources that belong to
the cluster, and the full set of variables pertaining to the proper
motion and CM subspaces. However, although normal densities are good
approximations to the distribution of data in the space of proper
motions, these do not model the distribution in the CMDs
satisfactorily . From stellar evolution theory we know that this
distribution follows a narrow set of isochrones that can be
represented as a multivariate curve in the colour-magnitude subspace.
The mixture-of-Gaussians likelihood model requires three components
(as assessed by the BIC) to reproduce the curvilinear distribution,
and the resulting covariances are too large to accommodate the
curvilinear nature of the distribution. This results in conspicuously
contaminated membership lists, and therefore we do not describe this
set of models further.

We obtained a curvilinear likelihood model by fitting a principal curve
\citep{hastie:principal} to the list of members. Principal curves
analysis is one of the various extensions of principal components
analysis to non-linear probability distributions that aims at
minimising the sum of square deviations in all variables in
representation space. This is accomplished with an iterative scheme
whereby the current version of the curve (initially the first
principal component) is modified in two steps: projection and
conditional expectation. The principal curve is represented by a set
of points $f_l$ (with $l=1,2,...,n$) in representation space that can
be interpolated with various models of different degrees of
complexity. Given our typical densities, a linear interpolation scheme
is more than sufficient. This piecewise linear interpolation defines a
parametrisation of the curve that can be thought of as a length along
the curve, measured from one of the extremes. We defined
$s(\lambda)$ as this parametrisation of the curve.

The projection step finds the point in the curve closest (in the
Euclidean sense) to a given observation $\vec{m}_{CM,l}$, and assigns
a coordinate value $\lambda_l$ to it. The conditional expectation step
redefines the set of points $\{f_l, l=1,2,...,n\}$ used for
interpolation as the expected value of colour-magnitude observations
with $\lambda=\lambda_l$:

\begin{equation}
f_l=E({\vec{x}_{CM}}|\lambda=\lambda_l),
\end{equation}

where $x_{CM}$ is the random variable from which the observations
$m_{CM}$ are drawn.

Since typically each observation projects onto a different value of
$\lambda$, some kind of smoothing scheme is needed in the expectation
step. We used a smoothing spline \citep[see
  e.g.,][]{hastie1990generalized} of $df$ degrees of freedom, where
$df$ depends on the representation space. This number of degrees of
freedom is the minimum necessary to model the distribution in the
various two-dimensional projections without conspicuous
biases. Several tests showed that principal curves obtained in a range
of degrees of freedom are barely distinguishable from one another and
produce the same candidate lists. When the iterations converged, the
same smoothing scheme was used to define the variance $\Sigma(\lambda)$
along the curve.

The final likelihood-model in the CM subspace is defined as

\begin{equation}
p(\vec{m}_{CM}|{\cal C}_2)=p(\lambda)\cdot p({\vec{m}_{CM}}|\lambda),
\end{equation}

where $p(\lambda)$ is interpolated from the empirical kernel density
estimate of sources per ${\rm d}\lambda$, and
$p({\vec{m}_{CM}}|\lambda)$ is modelled with a multivariate normal
distribution, the covariance of which is again interpolated from the
empirical distribution of covariances estimated from sets of sources
within intervals of $\Delta\lambda$. This effectively overestimates
the true covariance since the estimated values respond not only to
the natural spread of the isochrone, but also to the measurement
errors.

We found clear evidence of an isochrone parallel to (and brighter
than) the main cluster component. This is interpreted as the result of
the presence of equal mass binaries in the cluster sample. In general,
the spread of the sequence is due to the combined effects of
multiplicity (at all possible luminosity ratios), intrinsic
variability, and rotation. We only modelled equal mass binaries with a
second likelihood-component identical to the cluster principal curve,
but displaced in the magnitude space by 0.75 magnitudes. The relative
prior probabilities for the two principal curves were set {\em ad hoc}
to 0.8 and 0.2 for the main sequence and the sequence of binaries,
respectively. These values were not updated during the iterations
because the binarity of the sources is not amongst the set of inferred
parameters, but also because this fraction varies with stellar
luminosity. We checked the consistency of this choice in view of the
resulting membership lists in Sect. \ref{sect:sensitivity}, and
discuss alternatives in Sect. \ref{sect:lim-concl}.

To prevent binaries from biasing the principal curve estimation, we
fitted it in two stages. In the first stage, we computed the principal
curve for the complete list of members in each iteration of the EM
algorithm. This fit is expected to be biased towards brighter
magnitudes due to the presence of binaries. We subsequently filtered
out sources that are brighter in all magnitudes of the representation
space than the closest (in the Euclidean sense) point in the principal
curve. We also removed points with Mahalanobis distances from this
closest point greater than 15 (a somewhat arbitrary but conservative
value that only removes clear outliers). We then recomputed the
principal curve with the filtered data set.

\begin{figure}[thb]
  \centering 
  \includegraphics[scale=0.25]{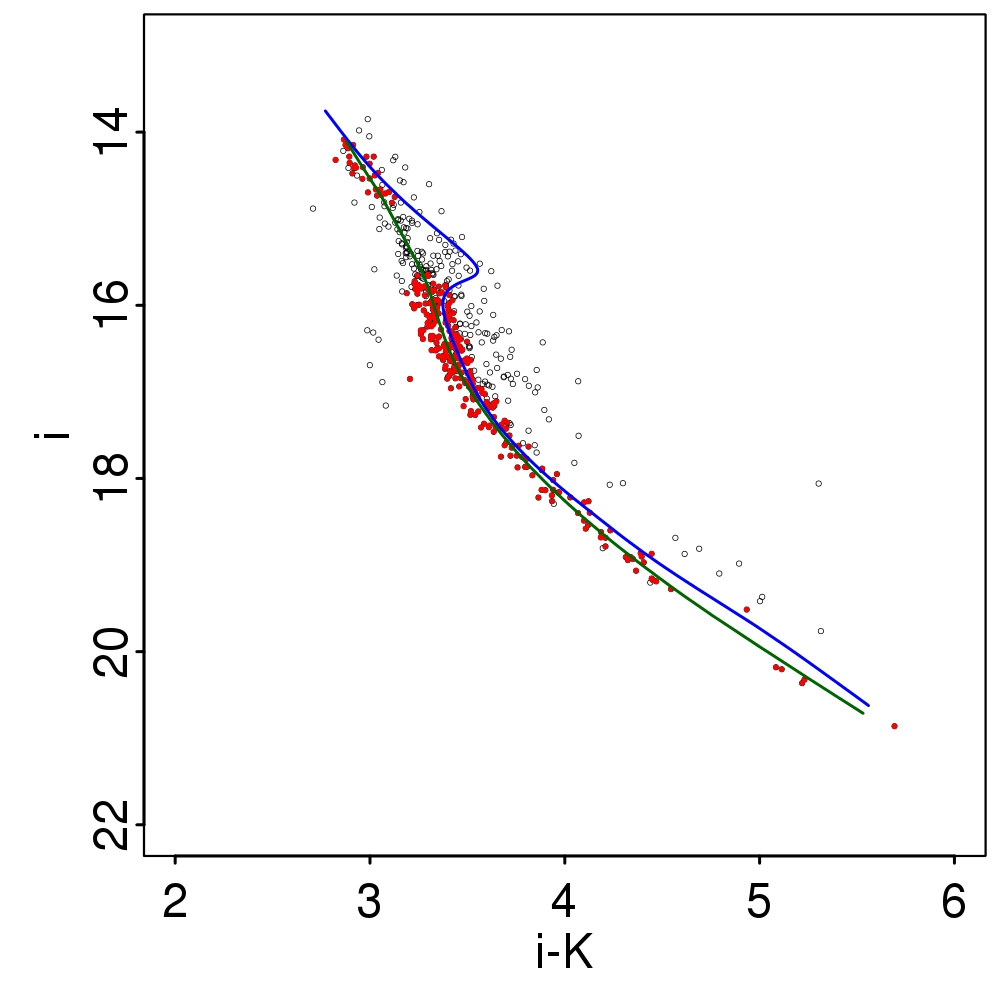}
  \caption{\label{firstPC} Principal curve fits to the initial
    reference set (blue line) and to the subset of sources with
    all magnitudes fainter than its closest point in the first principal
    curve (green line). This subset of points is represented in red.}
\end{figure}

Figure \ref{firstPC} shows these two consecutive fits to the initial
reference data. The blue line corresponds to the first principal
component fit to the initial reference set (with outliers removed),
while the green line represents the second fit to the subset
represented as red circles.

As mentioned above, we maintain the mixture-of-Gaussians model, and
the model complexity selection based on the BIC for the subspace of
proper motions of cluster members.

\subsection{Iterative refinement of the initial reference set}

In the previous paragraphs we have described the derivation of the
likelihood models from the initial reference set defined by the
membership list in \cite{2007ApJS..172..663S}. Close examination of
the membership probabilities derived from Eq. \ref{bayes.th} and these
likelihood models reveals the presence of sources in the initial
reference set with low membership probabilities and {\em vice
  versa}. Therefore, and for the sake of self-consistency, we adopted an
iterative procedure, based again on the EM algorithm, whereby we
derived likelihood models in iteration $l$ from the membership lists
calculated in iteration $l-1$ (maximisation step). Then, we used these
likelihood models to infer membership probabilities (expectation
step), and proceeded to the next iteration, $l+1$.

The first iteration proceeded as described in the previous sections,
with the data set $\mathcal{D}$ defined by the set of sources observed
in all the variables in representation space, and with the class
labels ${\cal C}$ taken from \cite{2007ApJS..172..663S}. Let
$p_0({\cal D}|{\cal C}_i)$ denote the initial likelihood models
derived from the initial reference set. In subsequent iterations, the
regularised maximum-likelihood model, together with
Eq. \ref{bayes.th}, were used to infer posterior membership
probabilities (expectation step). Since the two models by definition
comprise all possibilities, we can compute the posterior probability
in iteration $l$ as

\begin{equation}
p_l({\cal C}_i|{\cal D}) = \frac{p_{l-1}({\cal D}|{\cal C}_i)\cdot p_{l-1}({\cal
    C}_i)}{\sum_{i=1}^2p_{l-1}({\cal D}|{\cal C}_i)\cdot p_{l-1}({\cal C}_{i})}.
\end{equation}

We took into account the measurement uncertainties in the likelihood
term as follows:

\begin{equation}
\label{uncertainties}
p_{l-1}(\vec{m}_{\rm measured}|{\cal C}_i)= \int p_{l-1}(\vec{m}|{\cal C}_i)\cdot
p_{l-1}(\vec{m}_{measured}|\vec{m}) {\rm d}\vec{m}
\end{equation}

where $\vec{m}_{measured}$ represents the measured values. We assumed
independent Gaussian uncertainties in all measured variables (proper
motions and apparent magnitudes) and thus,

\begin{equation}
p_{l-1}(\vec{m}_{measured}|\vec{m})=\mathcal{N}_{\vec{m},\Sigma}(\vec{m}_{measured}),
\end{equation}

where $\Sigma$ represents the non-diagonal covariance matrix of the
variables in representation space. The matrix elements were derived
following the usual rules for uncertainty propagation of variables
that are linear combinations of the original ones (the colour
indices).

We then redefined the membership lists according to these posterior
probabilities. We used two heuristic thresholds $p_{in}$ and $p_{out}$,
such that members with posterior membership probabilities below
$p_{out}$ were removed from the training set of cluster members, while
non-members with posterior membership probabilities above $p_{in}$ were
included in it. We have tested three values of
$p_{in}=\{0.9545, 0.975, 0.9975\}$ (probability values corresponding
to the range from 2$\sigma$ to 3$\sigma$ significance), and set
$p_{out}=0.5$. This represents criteria that are very restrictive for
the admission of sources not previously classified as members in the
literature, while somewhat conservative for the rejection of members.

With these updated membership lists we derived new likelihood models
$p_l({\cal D}|{\cal C}_i)$ and prior probabilities (the maximisation
step), and iterated the EM scheme until the membership lists remained
unchanged.

We stress that the implicit extrapolation of the principal curve
beyond the range of magnitudes and colours covered by the initial
reference set, carried out during the EM iterations, is not blind, but
constrained to be based on sources with high membership probabilities,
and thus is highly consistent with the distribution in the proper motion
space.

When the iterations converged to a stable solution (defined by no
change in membership for any source in two consecutive iterations), we
computed membership probabilities for incomplete sources (i.e.,
sources with missing data) by again applying Bayes' theorem, with a
likelihood model obtained by marginalising the complete
maximum-likelihood model derived in the previous section over the
missing variables:

\begin{equation}
p(\vec{m}_{measured}|\mathcal{C}_i)=\int p(\vec{m}|\mathcal{C}_i) \cdot {\rm d}\vec{m}_{missing},
\label{eq:missing}
\end{equation}

where $\vec{m}=\{\vec{m}_{measured},\vec{m}_{missing}\}$, and
$\vec{m}_{measured}$ and $\vec{m}_{missing}$ are the subsets of
$\vec{m}$ that are measured and missing, respectively. Since the model
for $p(\vec{m}|\mathcal{C}_i)$ is a mixture of Gaussians, the result
of the integral is again a mixture of Gaussians with means and
covariance matrices that correspond to the projection of the original
ones onto the subspace of measured variables. Uncertainties are
handled in the same way as described in Eq. \ref{uncertainties}.

\section{Representation space}
\label{sect:repspace}

It is well known that not all variables are equally informative for
the determination of membership probabilities. As a matter of fact,
apparent magnitudes alone may be insufficient for the selection of
member candidates, and it is common practice to define hard boundaries
in CMDs \citep[see e.g.][]{2012MNRAS.422.1495L}. Including
non-informative variables may also dilute the discriminative power of
relevant ones, thus leading to poor performances. We explored
several representation spaces and assessed their relative
merits.

As shown in Table \ref{nrofmeas}, 89\% and 79\% of the sources do not
have measurements in the u and g bands, respectively. We excluded these
two bands to avoid creating initial models of the member and field
samples composed exclusively of bright, relatively hot objects.

We assessed a number of representation spaces summarised in Table
\ref{repspaces}. 

\begin{table*}
\caption{Representation spaces evaluated in this work.}
\label{repspaces}
\centering       
\begin{tabular}{l  l}
\hline                                
Experiment & Representation space \\
\hline\hline
Apparent Magnitudes (AM) & $\mu_{\alpha}$ $\mu_{\delta}$, r, i, J, H, K, Y, Z\\
\hline                                    
Colour indices 1 (CI-1) &  $\mu_{\alpha}$ $\mu_{\delta}$, (r-i), (i-Z), (Z-Y), (Y-J), (J-H), (H-K)\\
\hline                                
Colour indices 2 (CI-2) &  $\mu_{\alpha}$ $\mu_{\delta}$, (i-J), (i-K), (Y-J), (Z-J), (r-H), (r-K)\\
\hline                                
r + CI-2 &  $\mu_{\alpha}$ $\mu_{\delta}$, r, (i-J), (i-K), (Y-J), (Z-J), (r-H), (r-K)\\
\hline                                
i + CI-2 &  $\mu_{\alpha}$ $\mu_{\delta}$, i, (i-J), (i-K), (Y-J), (Z-J), (r-H), (r-K)\\
\hline                                
K + CI-2 &  $\mu_{\alpha}$ $\mu_{\delta}$, K, (i-J), (i-K), (Y-J), (Z-J), (r-H), (r-K)\\
\hline                                
RF-1 & $\mu_{\alpha}$ $\mu_{\delta}$, r, J, H, K, (i-J), (r-K)\\
\hline                                
RF-2 & $\mu_{\alpha}$ $\mu_{\delta}$, J, H, K, (i-K), (r-K), (Y-J)\\
\hline                                
RF-3 & $\mu_{\alpha}$ $\mu_{\delta}$, J, H, K, Y, (i-Z), (r-Y), (r-i)\\
\hline                                
\end{tabular}
\end{table*}

The first set of variables corresponds to the set of apparent
magnitudes; the second set corresponds to the set of colour indices
defined by two consecutive spectral bands; the third set corresponds
to a heuristic set defined intuitively as relevant for the separation
of the two classes; the fourth, fifth, and sixth sets correspond to
the heuristic feature set with one apparent magnitude added to it (r,
i, and K, respectively); the seventh and eighth feature sets were
obtained by performing an importance analysis with random forests (see
below for a more detailed explanation) induced from the set of
features defined by the heuristic set of colour indices plus the set
of apparent magnitudes; finally, the ninth feature set was again
derived from an importance analysis with random forests, but induced
now from the entire set of apparent magnitudes and all potential
colour indices.

The variable sets RF-1 and RF-2 in Table \ref{repspaces} are defined
by the most important variables, with importance measured by the mean
decrease in accuracy, and by the mean node impurity obtained for
out-of-bag (OOB) samples and a random forest classifier
\citep{Breiman-RF}. 

A random forest is a relatively large number of classification trees.
A classification tree, in turn, is a tree-like decision graph that
predicts the class of an input case (in our context, whether a source
belongs to the cluster or field subsets) by successively applying
tests on subsets of the representation space variables. The tests on
the variables take the form of conjunctions of conditionals, for
example, $m_i >= \zeta_i$ AND $m_j < \zeta_j$, where $\zeta$ acts as a
decision boundary, and we omitted the distinction between measured and
true values of the variables to simplify notation. These tests are
represented as nodes in the tree; each arc connecting nodes represents
possible outcomes of the test; and the leaf nodes represent class
assignments. Each classification tree in a random forest is
constructed using random subsets of representation space variables for
the tests performed in each of the tree nodes. A more detailed
explanation of the induction algorithm that constructs the tree from a
given training set is beyond the scope of this article. Descriptions
of the methodology used here can be consulted for instance in
\cite{Murphy2012Machine}.

The decision boundaries in each node of each tree in the random forest
are inferred from bootstrap samples of the initial reference data set,
from which a set of examples was separated for testing purposes
(the OOB sample). By randomly shuffling the values of a given
representation space variable and subsequently quantifying the
classification performance degradation, we gauged the relative
importance of that variable. We expect that shuffling of the values of
irrelevant (non-informative) variables will result in little or no
performance degradation. In the case of RF-1 and RF-2, the random
forests were induced, as mentioned above, by using a training set
comprising all apparent magnitudes and the set of colour indices
defined heuristically (CI-2), while RF-3 was defined by the random
forest induced from a training set that includes all apparent
magnitudes and all potential colour indices. In all cases, proper
motions, magnitudes, or colour indices were added in decreasing order of
importance (starting from an empty set) until the next variable is a
linear combination of the current set of variables.  The total number
of variables used in the various models is thus limited by the
requirement that the covariance matrices in the mixture-of-Gaussians
have to be invertible. Proper motions appear selected in the top
places of the ranking in all cases.

\begin{table*}
\caption{Mean decrease in accuracy and node impurity of random forests
  trained with the heuristic set (columns 2 and 3) or the complete set
  (columns 5 and 6) of proper motions, magnitudes, and colour
  indices.}
\label{RFFS}
\centering       
\begin{tabular}{l l l | l l l}
\hline                                
Variable & \multicolumn{2}{c}{Heuristic set: Mean decrease in} & Variable & \multicolumn{2}{c}{Full set: Mean decrease in} \\
& classification  & node     & & classification  & node  \\
& accuracy        & impurity & & accuracy        & impurity \\
\hline\hline    
$\mu_{\delta}$ &   0.0021   &      530.6 &  $\mu_{\delta}$ & 0.0016 & 369.6\\
\hline                                
$\mu_{\alpha}$ &   0.0015   &      199.4 &  $\mu_{\alpha}$ & 0.0010 & 130.1\\
\hline                                
r-K          &   0.0014   &      156.0  & K    & 0.0013 & 114.8 \\
\hline                                
K            &   0.0024   &      127.1  & J    & 0.0016 & 98.0  \\
\hline                                
i-K          &   0.0008   &      102.0  & H    & 0.0011 & 85.7\\
\hline                                
H            &   0.0012   &       80.1  & r-Z  & 0.0008 & 82.0\\
\hline                                
Y-J          &   0.0006   &       77.9  & Y    & 0.0014 & 70.6\\
\hline                                
J            &   0.0017   &       77.3  & r-Y  & 0.0007 & 60.2 \\
\hline                                
r-H          &   0.0005   &       48.7  & r-H  & 0.0006 & 57.5\\
\hline                                
Y            &   0.0009   &       41.8  & r-J  & 0.0006 & 55.4  \\
\hline                                
Z-J          &   0.0002   &       27.3  & Z    & 0.0011 & 47.0\\
\hline                                
Z            &   0.0007   &       21.6  & i-Y  & 0.0004 & 44.5\\
\hline                                
i            &   0.0009   &       18.8  & r-i  & 0.0004 & 43.1\\
\hline                                
r            &   0.0011   &       17.5  & r-K  & 0.0004 & 42.9\\
\hline                                
i-J          &   0.0013   &       16.0  & i-Z  & 0.0004 & 36.5\\
\hline                                
\end{tabular}
\end{table*}

We visually checked the results (final reference sets and membership
probability distributions for the entire DANCe data set) of the
algorithms for each representation space in Table \ref{repspaces}. In
the current work we only show detailed results for three
representation spaces (AM, RF-2, and RF-3) selected on the basis
of three criteria: higher exhaustivity (we preferred representation
spaces that produced isochrones that reached fainter magnitudes),
lower contamination rates (we avoided representation spaces that
result in conspicuously contaminated membership lists), and
conciseness (we did not include representation spaces that produced the
same reference set and thus, membership probabilities as one already
selected).

\section{Results}
\label{sect:results} 

In this section we discuss the results obtained for the RF-2
representation space and $p_{in}=0.975$, for the model described in
Sect. \ref{sect:rml} (that is, a mixture-of-Gaussians model for the
field sources in the full RF-2 representation space, and a mixed model
for cluster sources composed of a mixture-of-Gaussians in the proper
motions subspace and a curvilinear model in the CM subspace). In
Sect. \ref{sect:sensitivity} we briefly discuss the sensitivity of
these results as we change $p_{in}$ and the representation space. In
particular, we compare the results discussed in this section with
those obtained for the RF-3 representation space in more detail.

We carried out the iterative scheme described above in two stages. In
the first stage we used the full representation space, while the second
stage was performed in a reduced subspace without variables (colour
indices) derived from the r band. This is because the number of
complete sources in the reduced subspace is significantly larger that
in the original subspace, as can be deduced from Table \ref{nrofmeas},
especially for the reddest sources. 

Since there are in fact known member candidates from the literature
that are complete in the reduced feature subspace, but incomplete in
the original RF-2 set (that is, members detected in all bands used in
the RF-2 set except in the r band), we added these sources to the
membership list obtained in the full feature space. This approach has
the disadvantage that the lack of constraints for the r magnitude
allows including of member candidates far apart from the
defining isochrone in the full space if the large distance is only
due to discrepancies in that magnitude.

Using the RF-2 representation space, the BIC selected 2 and 26 Gaussian components for the likelihood models of the cluster
(proper motion subspace) and field populations, (full RF-2
representation space) respectively.

\begin{figure*}[thb]
  \centering
  \subfigure[]{\label{pm-st3}
  \includegraphics[scale=0.12]{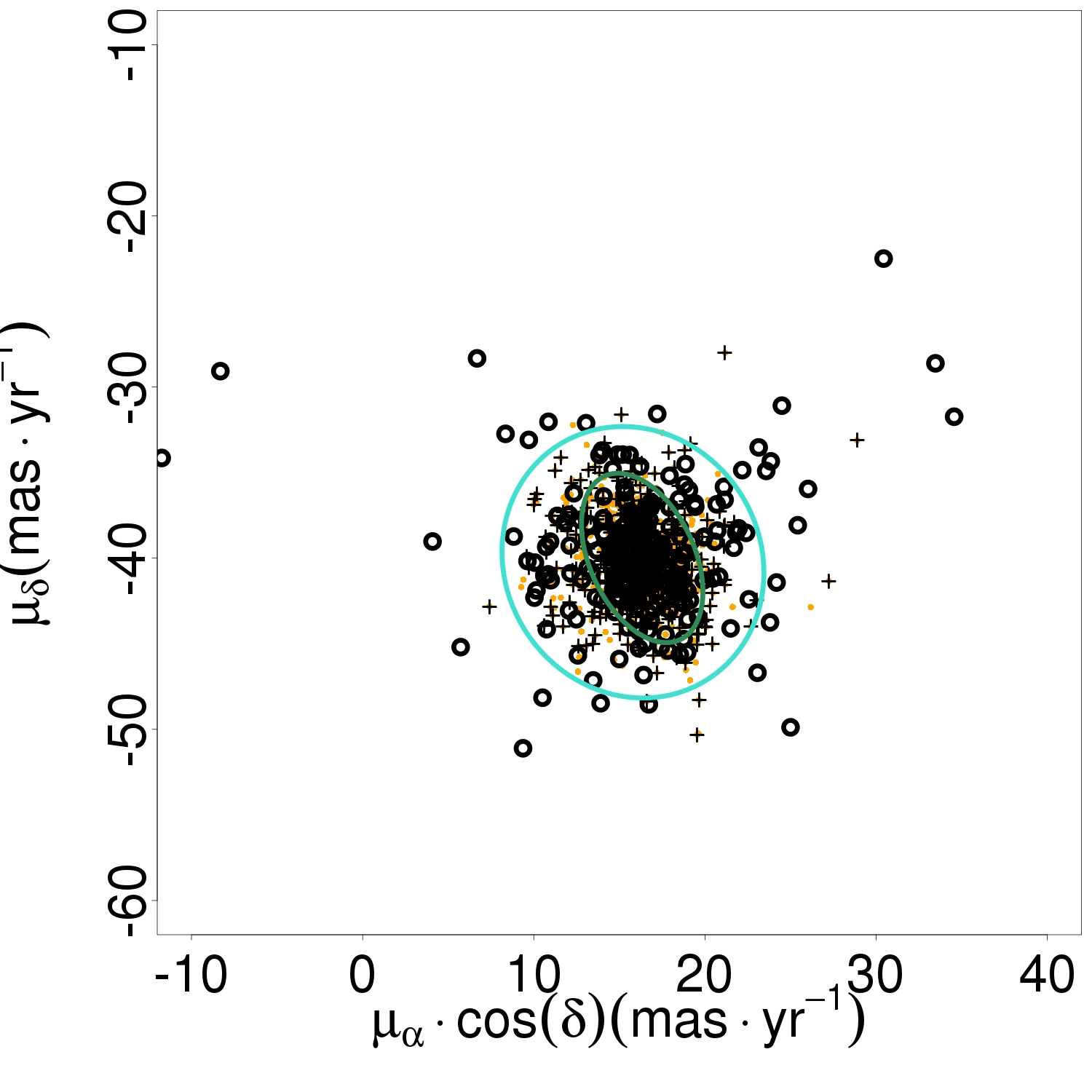}}
  \subfigure[]{\label{Kvsi-K-st3}
  \includegraphics[scale=0.12]{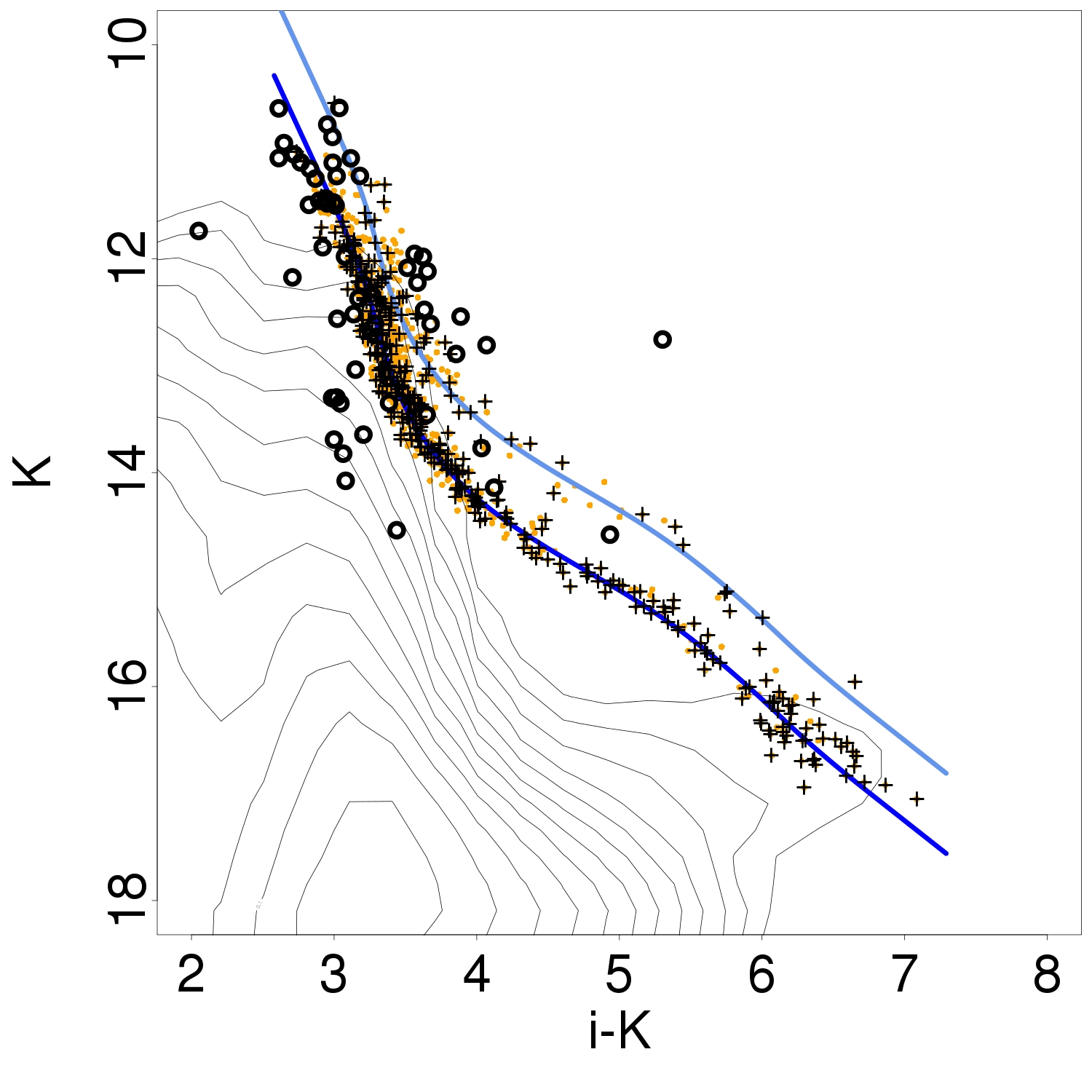}}
  \subfigure[]{\label{Kvsr-K-st3}
  \includegraphics[scale=0.12]{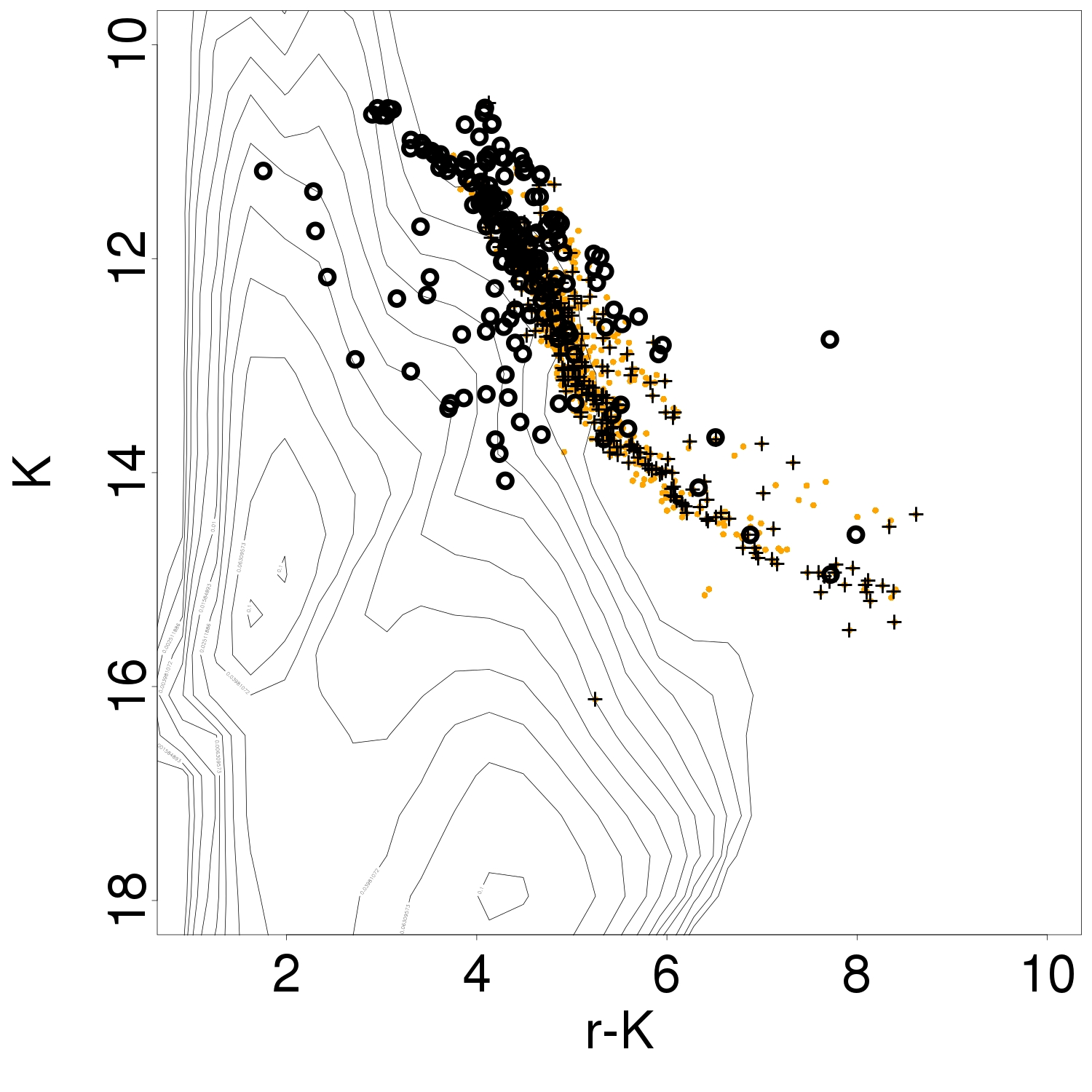}}
  \subfigure[]{\label{ivsi-K-st3}
  \includegraphics[scale=0.12]{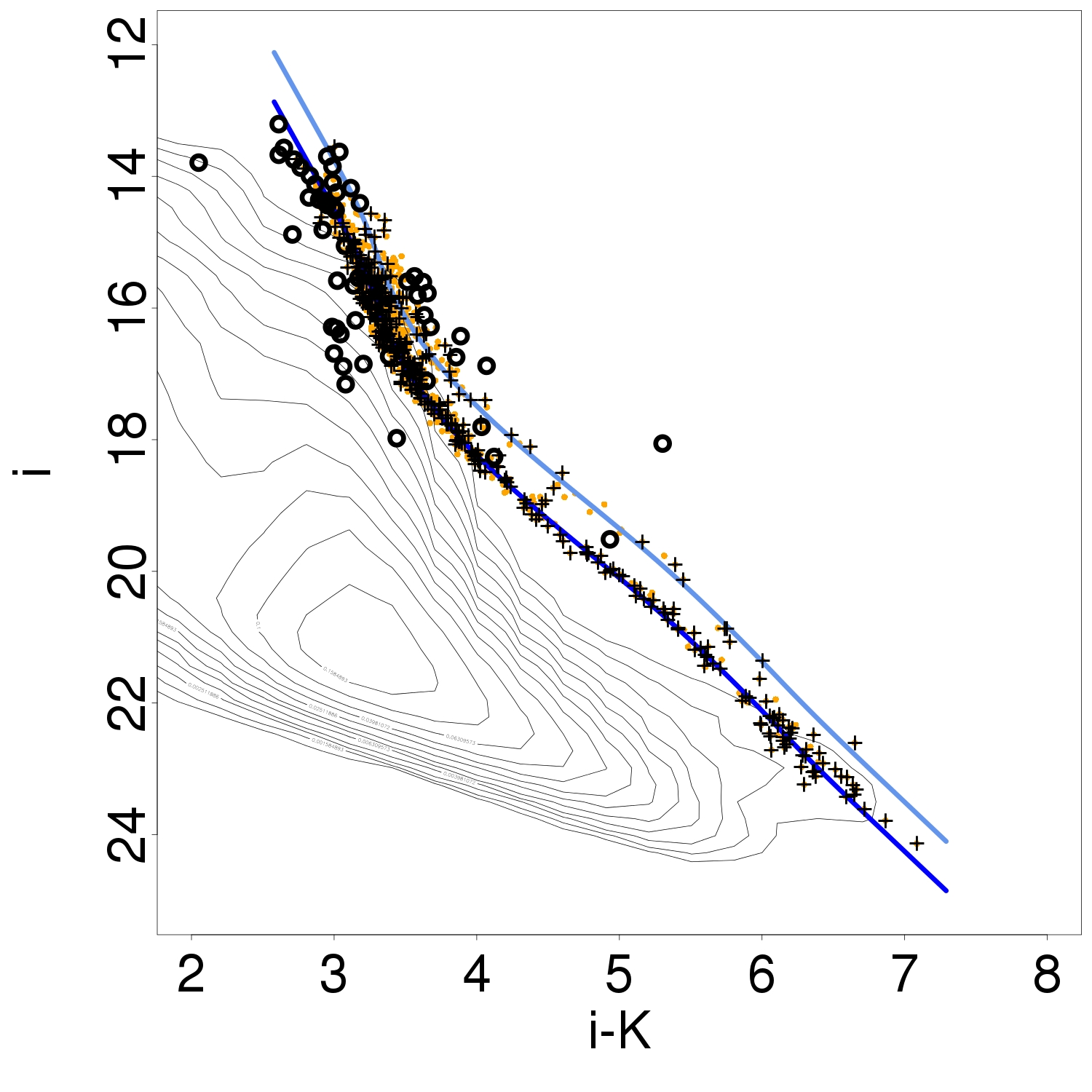}}
  \caption{\label{plot-st3} Distribution of the sources that define
    the combined curvilinear and Gaussian likelihood model (the final
    reference set after all EM iterations) in the same 2D
    projections used in Fig. \ref{data} (the VPD, and the (i-K)--K,
    (r-K)--K, and (i-K)--i CMDs). Black circles represent sources
      in the initial reference set that were removed during the EM
      iterations, and black crosses represent complete sources added
      in the iterative process. Orange dots represent sources in both
      the initial and final reference sets.}
\end{figure*}

Figure \ref{plot-st3} shows the final reference set of cluster members
(complete sources used to define the likelihood model) for this
representation space.  The contour lines represent the iso-density
lines computed using a Gaussian kernel density estimation obtained
from the entire data set. Figure \ref{pm-st3} shows the proper motion
diagram, while Figs. \ref{Kvsi-K-st3}, \ref{Kvsr-K-st3}, and
\ref{ivsi-K-st3} show the projections onto three CMDs. The coloured
ellipses in Fig. \ref{pm-st3} correspond to the 1-$\Sigma$ level of
the two Gaussian components in the proper motion subspace, obtained as
a result of the methodology described in the previous sections. The
fractional proportions of sources in each of these two Gaussian
components are $\pi_{2,1}=$ 0.62 (turquoise) and $\pi_{2,2}=$ 0.38
(green).  Sources included in the reference or training set during the
last phase of iterations (with missing r magnitudes) do not appear in
panel \ref{Kvsr-K-st3}. The contour lines in Fig. \ref{Kvsr-K-st3}
show a clear bimodality, produced by the combination in the DANCe data
set of two surveys with very different sensitivities.

The final number of sources in the set that defines the model
components is 886, of which 531 were in the initial reference set from
\cite{2007ApJS..172..663S}. The remaining 355 new sources are
marked with crosses in the various 2D projections in
Fig. \ref{plot-st3}. On the other hand, 222 members in the
initial membership list, were removed because of incompleteness, or
during the EM iterations. These are marked with empty circles. In
general, we see that the overall membership distribution in the
scatter plots is maintained with respect to the initial reference set,
except for the bright end of the K-(r-K) sequence, where the lack of
complete observations is the cause of the truncation. As we show in
the next section, this has a limited impact on the final list of
members for the full data set. Apart from these incomplete sources,
only some of the most conspicuous outliers are discarded after the EM
iterations.

The last stage of iterations in the reduced space (with the r
magnitude removed from the representation space) allows for a
definition set that reaches fainter magnitudes (i=24), illustrating
the importance and advantages of having homogeneous coverage and
sensitivity over the entire area of a survey.

\subsection{Membership probabilities for sources with missing measurements}
\label{missings-st3}

We subsequently use Eq. \ref{eq:missing} to derive membership
probabilities for sources with missing measurements in the RF-2
representation space. We define in general the list of cluster
  members as the set of sources with membership probabilities above
  $p_{min}$ (not to be confused with $p_{in}$, the probability
  threshold to include a complete source in the definition set).

\begin{figure*}[thb]
  \centering \subfigure[]{\label{all-pm-st3}
    \includegraphics[scale=0.12]{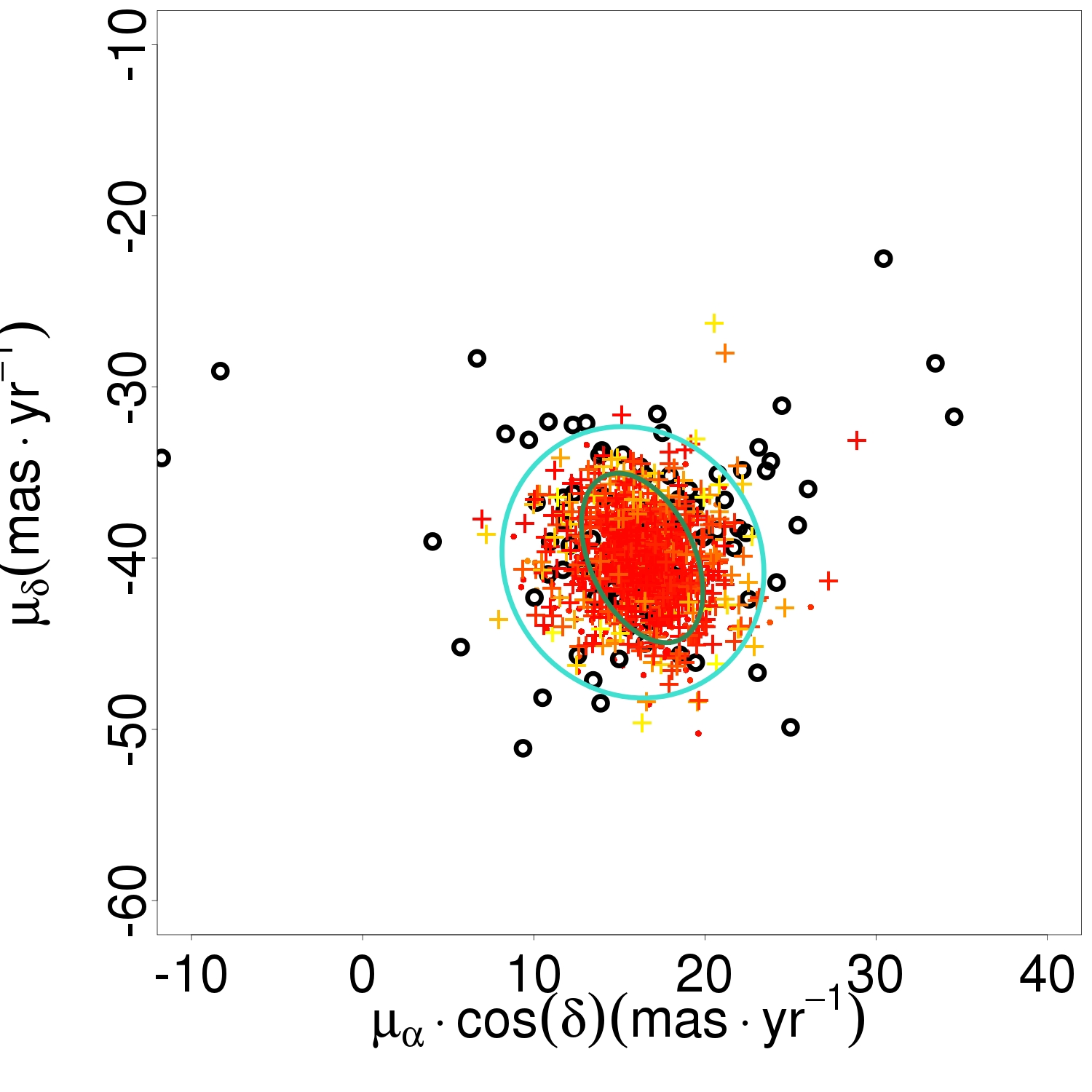}}
  \subfigure[]{\label{all-Kvsi-K-st3}
    \includegraphics[scale=0.12]{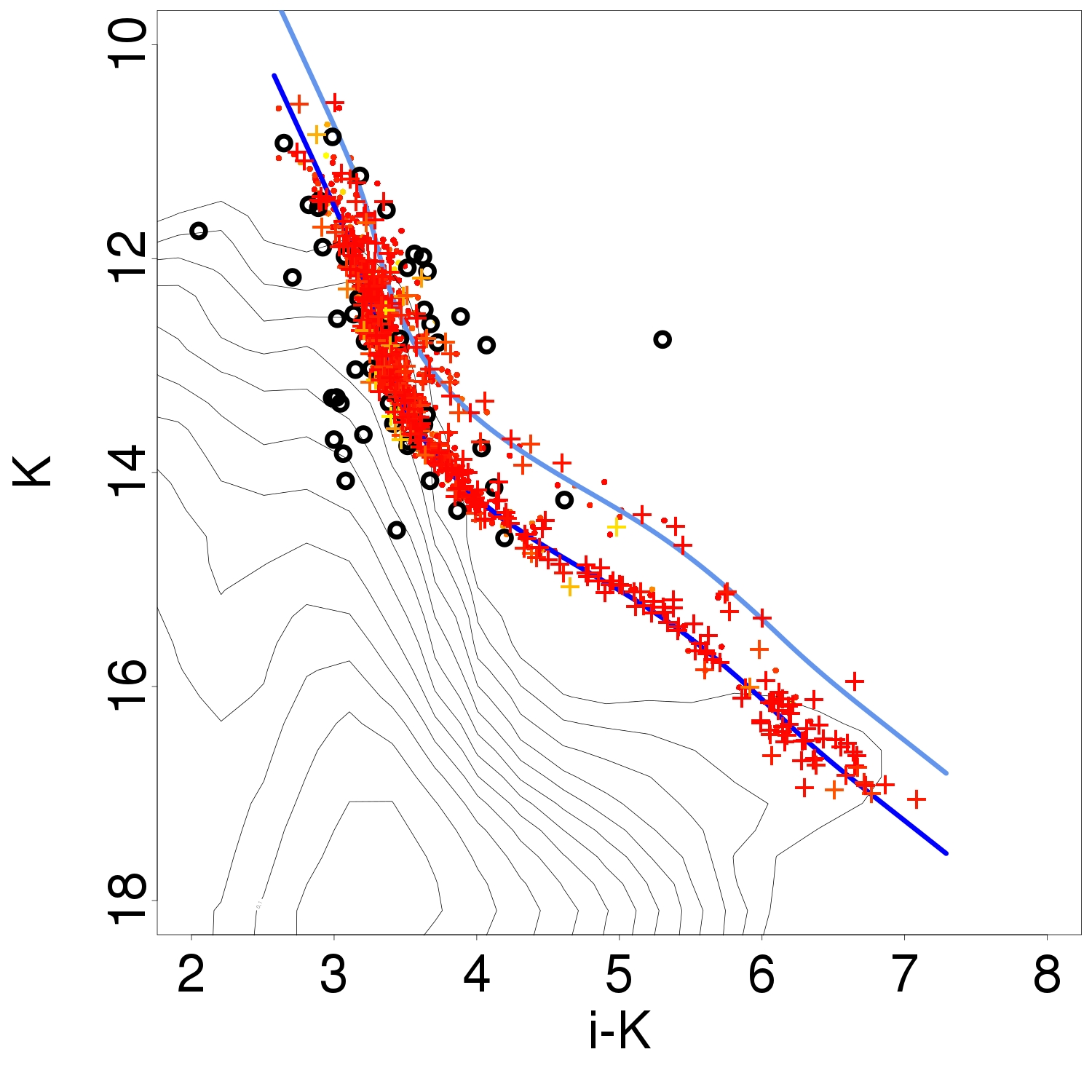}}
  \subfigure[]{\label{all-Kvsr-K-st3}
    \includegraphics[scale=0.12]{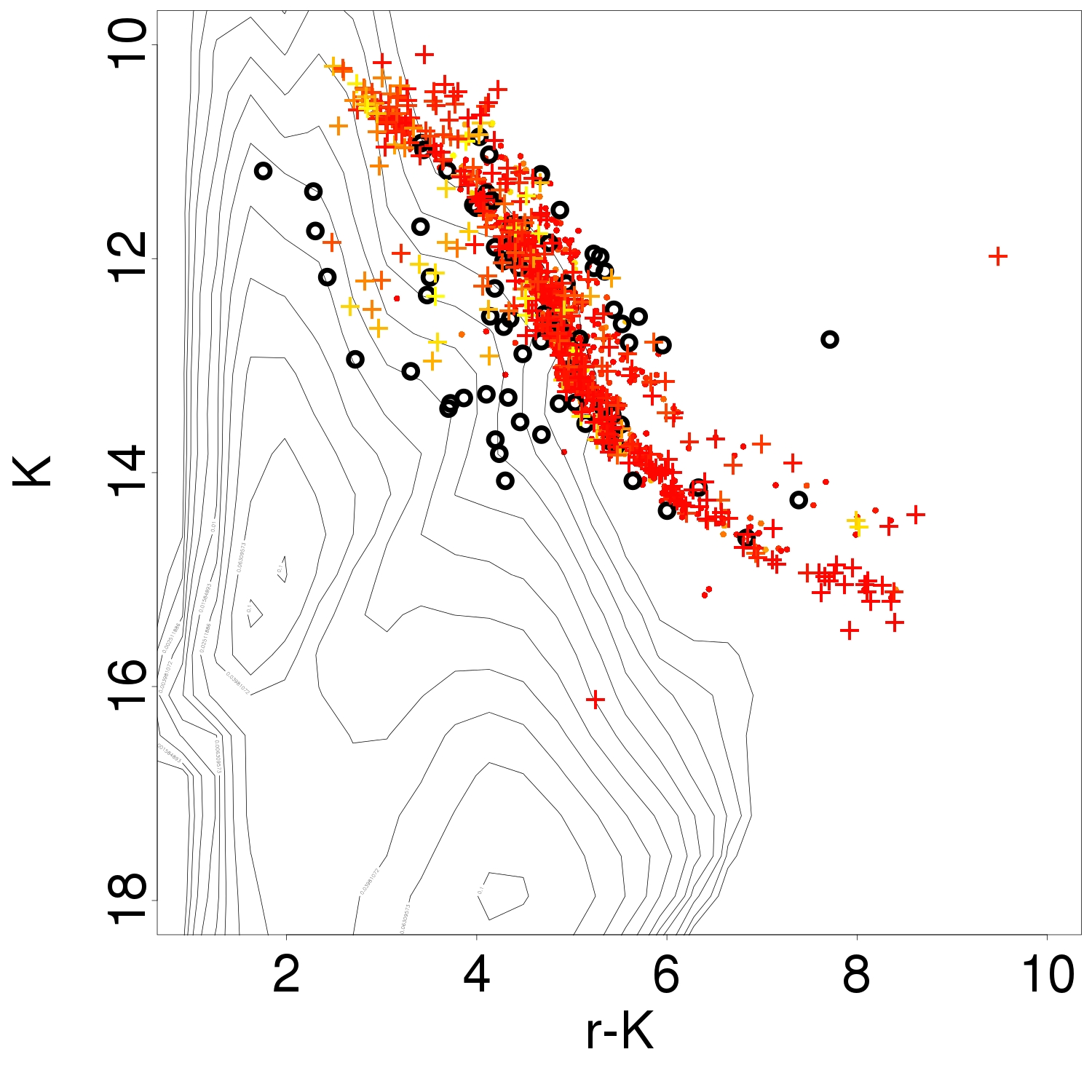}}
  \subfigure[]{\label{all-ivsi-K-st3}
    \includegraphics[scale=0.12]{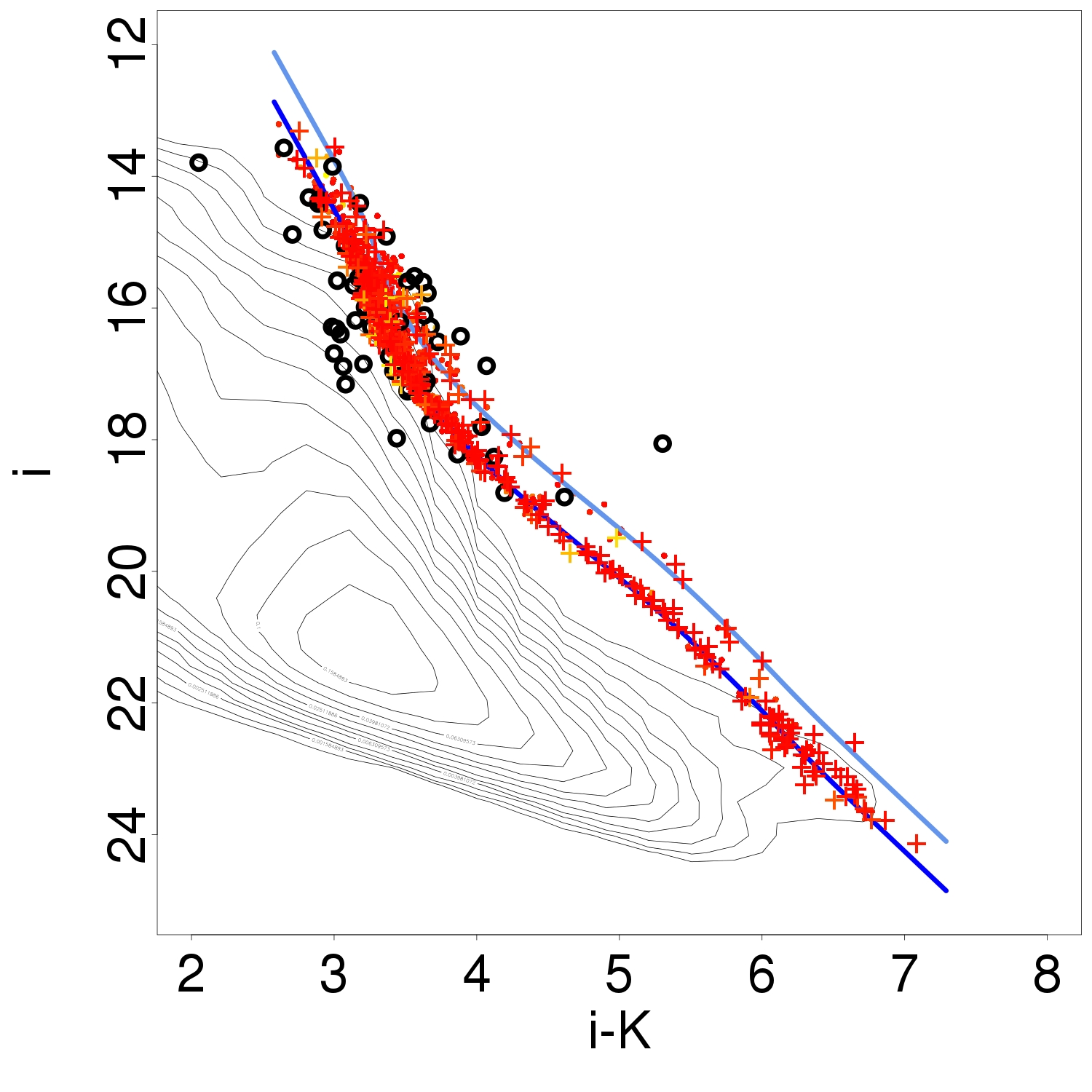}}
  \caption{\label{plot-missings-st3} DANCe catalogue sources
      (both complete and incomplete) with a membership probability
      above 0.975. Small dots represent sources in the original
      definition list; crosses represent member candidates not in the
      original definition list; black circles represent sources in the
      initial reference set with final membership probabilities below
      $p_{min}=0.975$. Contour lines trace the kernel density
    estimate obtained from the full data set. The colour code
    represents membership probability in a linear scale from yellow
    (p=0.975) to red (p=1).}
\end{figure*}

Figure \ref{plot-missings-st3} shows the proper motion
(\ref{all-pm-st3}) and CMDs (\ref{all-Kvsi-K-st3}
for K versus i-K, \ref{all-Kvsr-K-st3} for K versus r-K, and
\ref{all-ivsi-K-st3} for i versus i-K) of the candidate member list
for $p_{min}=0.975$. As in previous plots, crosses represent sources
with membership probabilities higher than $p_{min}$ not in the
initial reference set (thus, new members), black circles represent
sources in the initial reference set with membership probabilities
below $p_{min}$, and the membership probability is represented with a
linear colour scale between yellow ($p=0.975$) and red (p=1.0).

For a membership threshold $p_{min}$=0.975 the
resulting list of members includes 652 of the 757 sources in the
  original membership list from the literature, and 604 new sources
marked as plus signs in Fig. \ref{plot-missings-st3}. For a somewhat
more restrictive threshold $p_{min}$=0.9975, the membership list
contains 479 sources in the reference definition set, and 349 new
  members.

\begin{figure}[thb]
  \centering 
  \includegraphics[scale=0.12]{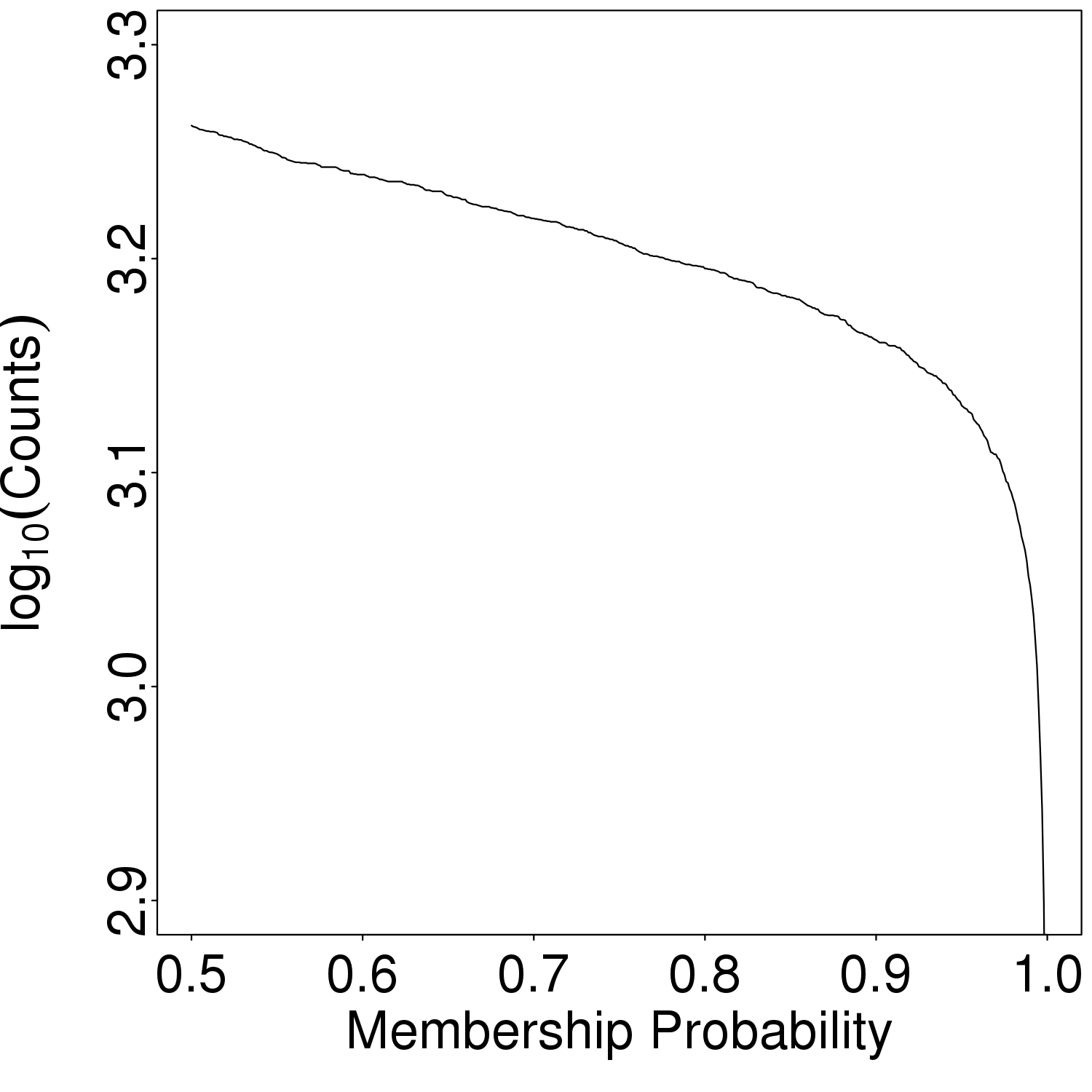}
  \caption{\label{cdf-pc} Cumulative number of sources (on a
    logarithmic scale) in the DANCe catalogue with membership
    probabilities (inferred with the principal-curve model and 
    $p_{in} = 0.975$, in the RF-2 representation space) above a
    given threshold.}
\end{figure}

It is evident from Fig. \ref{all-Kvsr-K-st3} that the lack of
constraints in the r magnitude results in a much more crowded K versus
r-K scatter plot, including a few clear outliers. Figure \ref{cdf-pc}
represents the cumulative number of sources in the DANCe catalogue
with membership probabilities above a given threshold.

Although not shown here for the sake of conciseness, the
colour-magnitude distributions obtained with a mixture-of-Gaussians
model for the entire representation space are remarkably similar in
shape to the results of the combined mixture-of-Gaussians plus
principal-curve model.

\subsection{Performance assessment with synthetic samples}

The probabilistic models (likelihoods) obtained from the EM iterations
can be used to generate artificial samples of sources, and the method
described in Sect. \ref{sect:rml} can then be applied to the synthetic
data set to determine whether it can recover the input set of members,
and assess the contamination level as a function of the threshold
adopted to define the membership. The validity of these assessments
will in turn depend on the validity of the inferred likelihoods.

We generated five samples of two million sources each, with the same
proportion of members as derived from the analysis of the DANCe data
set. Initially, the two million sources in each sample are generated
without missing values, and then some of the magnitudes were masked as
not available following the frequencies determined from the DANCe data
set. Finally, we assigned to each source in the synthetic catalogue
the measurement uncertainties of the closest (in the Euclidean sense)
DANCe source with the same pattern of missing variables. This means
that in the ideal case, we would have generated a new sample of
sources from the same probability distribution that generated the
DANCe data set. Of course, this would only be true if the model
inferred from the DANCe data set was exactly the probability
distribution that generated it.

Table \ref{tab-synth} summarises the evaluation of our algorithm on
the synthetic data set, expressed in terms of two quantities: the true
positive rate (TPR, the fraction of real members recovered by the
algorithm) and the contamination rate (CR, the fraction of the list of
member candidates drawn from the field density distribution). Each
quantity is accompanied by an estimate of the uncertainty derived from
the range of values observed in the five samples drawn from the
distribution. The uncertainty is defined as the radius of that
interval.

\begin{table*}

\caption{True positive rates and contamination rates for
  different values of the membership threshold. The uncertainty
  intervals correspond to the range of values (maximum-minimum)
  observed in the five random samples.}
\label{tab-synth}      
\centering                                      
\begin{tabular}{c c c c c c c c c c c c c c c c}          
\hline                                             
$p_{min}$ & 0.50 & 0.7 & 0.8  & 0.90 & 0.95  & 0.96 & 0.97 & 0.98 & 0.99  & 0.9975 \\
\hline
TPR (\%)  & 98.4$\pm$ 0.5 &  97.1$\pm$0.7 &  96.0$\pm$0.9 & 92.9$\pm$1.5 & 88.0$\pm$2.8 & 85.9$\pm$3.0 & 82.6$\pm$3.2 & 76.7$\pm$4.9 & 63.8$\pm$7.7 & 36.3$\pm$7.7 \\
CR (\%)   & 11.0$\pm$2.0 &  8.0 $\pm$1.5 &  6.6 $\pm$1.3 & 4.5$\pm$1.1 & 2.9$\pm$0.5  &  2.6$\pm$0.6 & 2.1 $\pm$0.5 & 1.6 $\pm$0.3 & 1.1 $\pm$0.3 & 0.4$\pm$0.4 \\
\hline\hline                        
\end{tabular}
\end{table*}

The ideal outcome of the experiment would be a classifier that
obtained a 100\% TPR without contamination. In that
respect, the point closest to this optimum in the curve defined by the
varying membership thresholds corresponds to $p=0.85$, TPR=92.9\% and
CR=4.5\%. We nevertheless considered that the optimum along the curve
cannot be defined uniquely because it is context dependent: some
research projects will demand very low contamination rates regardless
of the membership list completeness, while others will have opposite
requirements. We therefore opted for providing the full list of
posterior probabilities and let the catalogue user define his/her
membership threshold.

\section{Sensitivity analysis}
\label{sect:sensitivity}

In this section we analyse the dependence of the set of members
obtained on three particular choices: the probability threshold to
include a source in the set that defines the model ($p_{in}$), the
representation space, and the probability threshold for inclusion in
the set of cluster members ($p_{min}$).

Table \ref{sensitivity-tab} summarises the number of sources used to
define the model in the principal curves framework, for three
representation spaces (RF-2, RF-3 and the set of apparent magnitudes)
and admission probability thresholds ($p_{in}$=0.9545, 0.975 and
0.9975). The first three columns contain the definition set sizes at
the end of the last set of iterations (those in the reduced space,
without r related features) while the rightmost three columns show the
number of sources in the entire DANCe data set (comprising complete
and incomplete sources) with membership probabilities above 0.75 and
0.9975. When comparing the left and right numbers in Table
\ref{sensitivity-tab}, it has to be kept in mind that members in the
reference set are only removed from it in subsequent iterations if
their membership probabilities fall below 0.5.  According to Table
\ref{sensitivity-tab}, a tendency for smaller definition set sizes is
to be expected from higher admission thresholds, but the influence of
these changes on the final membership lists sizes is far from
straightforward.

\begin{figure*}[thb]
  \centering
  \includegraphics[width=\textwidth]{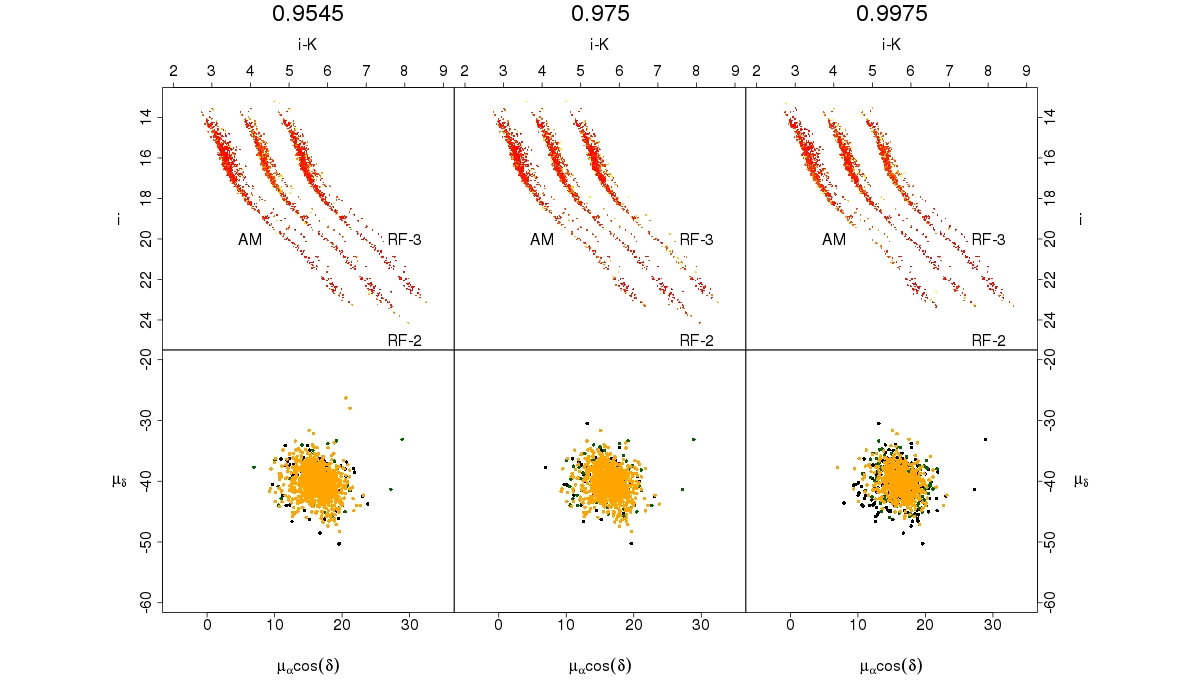}
  \caption{\label{sensitivity-fig} Vector point diagrams (bottom row)
    and (i-K)--i CMDs (top row) of sources in the
    DANCe catalogue (both complete and incomplete) with membership
    probabilities above 0.9975, for three values of the admission
    threshold $p_{in}$ to the definition set (0.9545 in the left
    column, 0.975 in the middle column, and 0.9975 in the right
    column). Each CMD shows the three cluster
    sequences obtained with the AM (left), RF-2 (middle), and RF-3
    (right) representation spaces. The colour code in the top row
    reflects the membership probability in a linear scale from yellow
    ($p$=0.9975) to red ($p$=1). In the VPD diagrams, we use black,
    green, and orange points to represent candidates obtained in the
    AM, RF-2, and RF-3 representation spaces.}
\end{figure*}

Figure \ref{sensitivity-fig} shows the distribution of sources
(complete or incomplete) with probabilities above 0.9975, after the
projection onto the reduced space (thus removing the r magnitude) in
the proper motion (bottom row) and i--(i-K) (top row) diagrams. The
cluster sequences in the i--(i-K) diagrams correspond, from left to
right, to the AM, RF-2, and RF-3 feature sets. The latter two are
displaced in the $x$-axis by one and two units, respectively, for the
sake of clarity. The colour code reflects the membership probability
on a linear scale, from yellow ($p=$0.9975) to red ($p$=1). The VPDs
in the bottom row show AM candidates as black points, RF-2 candidates
as green points, and RF-3 candidates as orange points. The three
feature spaces convey very similar cluster sequences except for a
larger scatter in the brighter segment of the AM feature set, and a
sequence reaching fainter sources in the RF-2 space with
$p_{in}=0.9545,0.975$.

\begin{table*}
\caption{Number of members in the final converged reference set (left)
    and in the DANCe catalogue (right).}
\label{sensitivity-tab}
\centering       
\begin{tabular}{l |l l l | l l l}
\hline                                
Admission & \multicolumn{3}{c}{Definition set} & \multicolumn{3}{c}{DANCe catalogue}\\
threshold & RF-2 & RF-3 & AM & RF-2 & RF-3 & AM \\
\hline\hline    
$p_{in}$=0.9545 (2$\sigma$)  & 911 & 882 & 910  & 1690:930 & 1543:916 & 1578:1014\\
\hline                                
$p_{in}$=0.975 (2.2$\sigma$) & 886 & 881 & 896  & 1612:828 & 1556:979 & 1567:1032\\
\hline                                
$p_{in}$=0.9975 (3$\sigma$)  & 827 & 810 & 827  & 1565:829 & 1474:769 & 1575:977 \\
\hline                                
\end{tabular}
\tablefoot{ The sizes of the definition set correspond to the final
  sets after convergence of all EM iterations.  The number of members
  in the DANCe catalogue is listed for membership probabilities higher
  than 0.75 and 0.9975 (separated by a colon). In both cases, the
  sizes are listed for three representation spaces (all magnitudes,
  RF-2, and RF-3) and probability thresholds (0.9545, 0.975, and
  0.9975).  }
\end{table*}

\begin{figure}[thb]
  \centering
  \includegraphics[scale=.25]{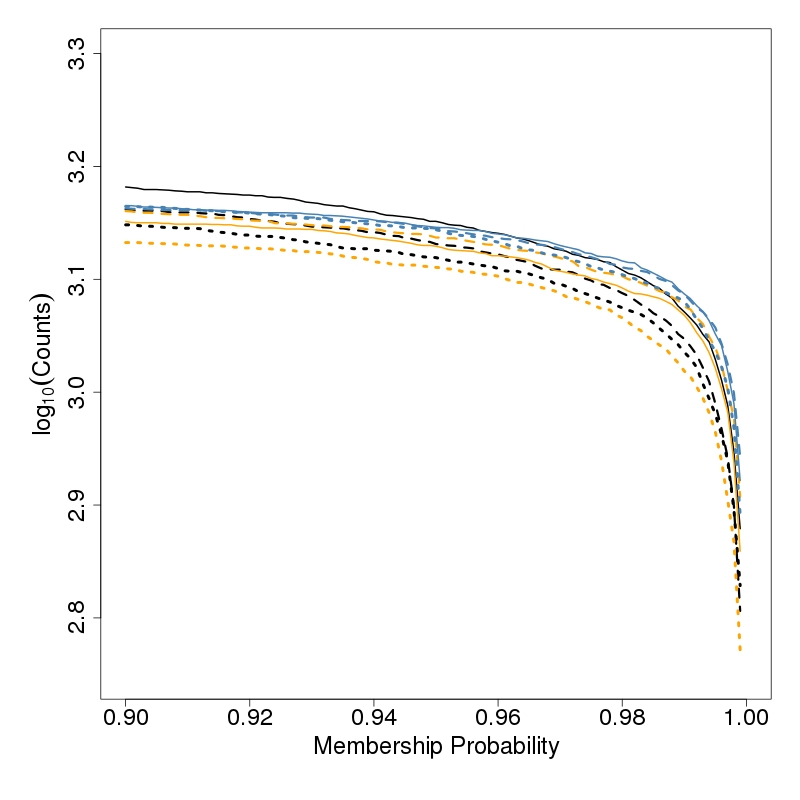}
  \caption{\label{cdf-pc-all}Cumulative number of sources in
    the DANCe catalogue with membership probabilities above a given
    value for the AM (blue), RF-2 (black), and RF-3 (orange) feature
    sets. Continuous lines represent models derived with
    $p_{in}=2\sigma$, long-dashed lines correspond to
    $p_{in}=2.2\sigma$, and short-dashed lines to $p_{in}=3\sigma$.}
\end{figure}

Figure \ref{cdf-pc-all} shows the cumulative number of sources (on a
logarithmic scale) with membership probabilities above a given
value. Continuous lines correspond to the 2$\sigma$ definition set
(black for RF-2, orange for RF-3, and blue for the AM feature set);
long-dashed lines correspond to the 2.2$\sigma$ threshold, and
short-dashed lines correspond to the 3$\sigma$ threshold (with the
same colour code in the three thresholds). We see that although all
feature sets result in relatively stable membership probabilities, the
AM set results are less affected by changes in the threshold for
inclusion in the reference set, especially as the membership threshold
is reduced. It also shows that more restrictive admission thresholds
$p_{in}$ result in shorter membership lists for a fixed probability,
except in the RF-2 space. In general, the relationship for a given
feature set between the $p_{in}$ threshold and the final membership
probabilities is not always straightforward. The intuition that a
higher threshold $p_{in}$ should result in lower membership
probabilities for the full data set does not always apply. This is
because the set of sources used for the model definition (that is, the
set of sources with measurements in the full set of features, referred
to in this work as complete sources) does not necessarily follow the
same probability density in the feature space as the full data set
(including the incomplete sources). Thus, a higher threshold may
result in changes in the $d$-dimensional model (2D Gaussian mixture
plus the principal curve) such that the new model passes through
regions with a higher or lower density of incomplete sources, or both,
in different curve segments. This problem would be solved if we used
all sources (complete and incomplete) to define the cluster
probabilistic model, as indicated in Sect. \ref{sect:lim-concl}.

\begin{figure*}[thb]
  \centering\subfigure[]{\label{psigmaam}
    \includegraphics[scale=0.17]{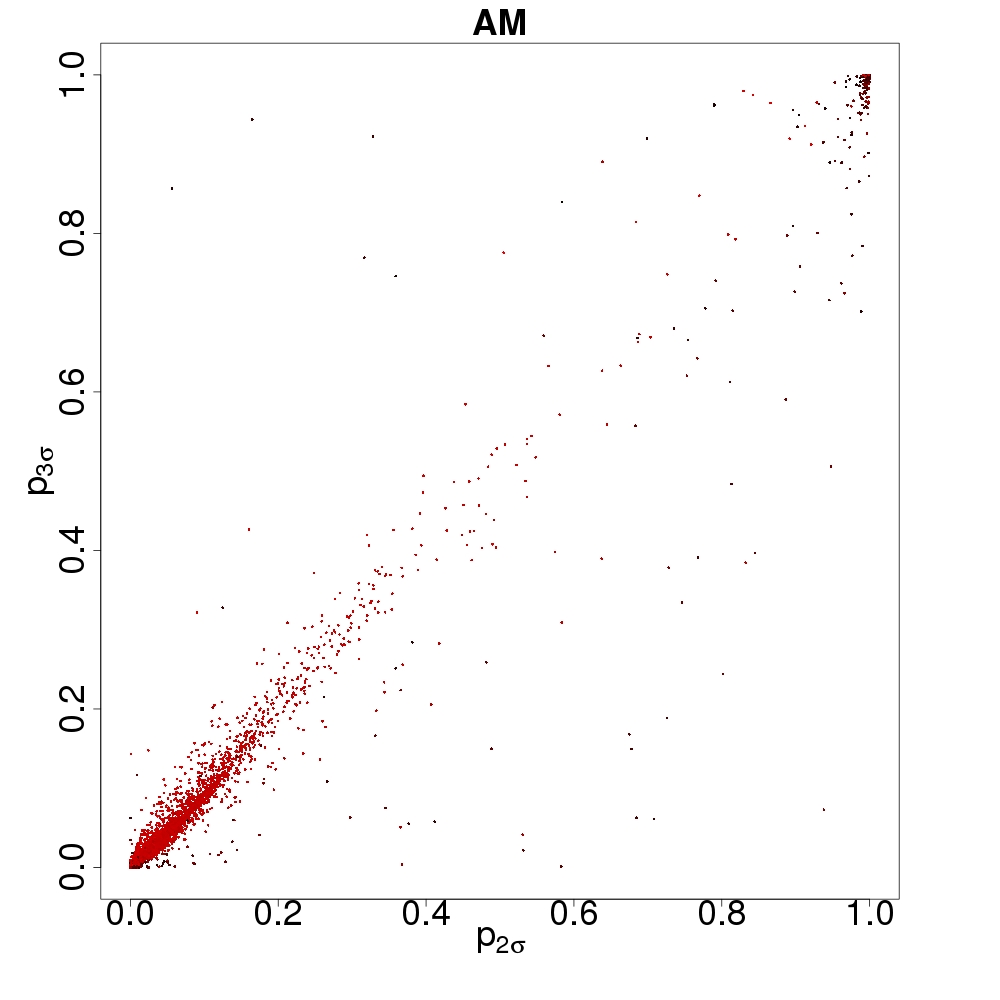}}
  \centering \subfigure[]{\label{psigmarf2}
    \includegraphics[scale=0.17]{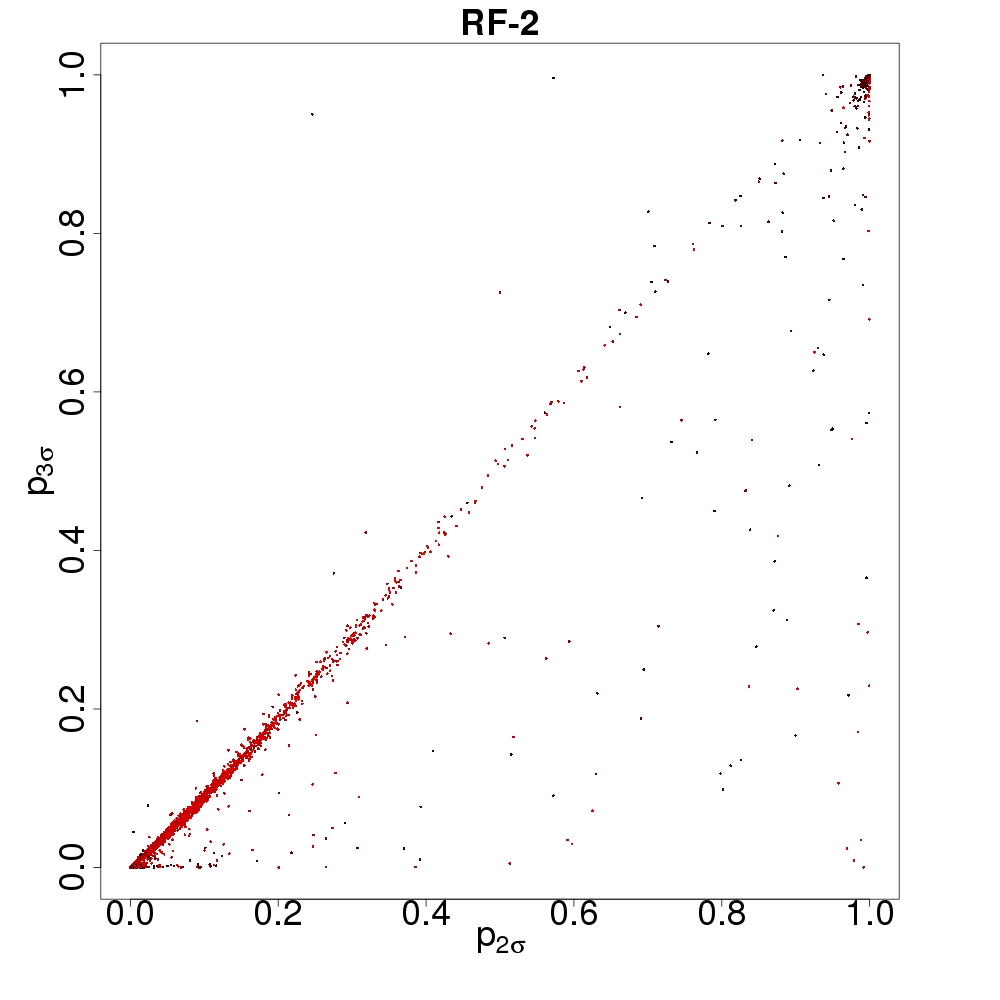}}
  \centering\subfigure[]{\label{psigmarf3}
    \includegraphics[scale=0.17]{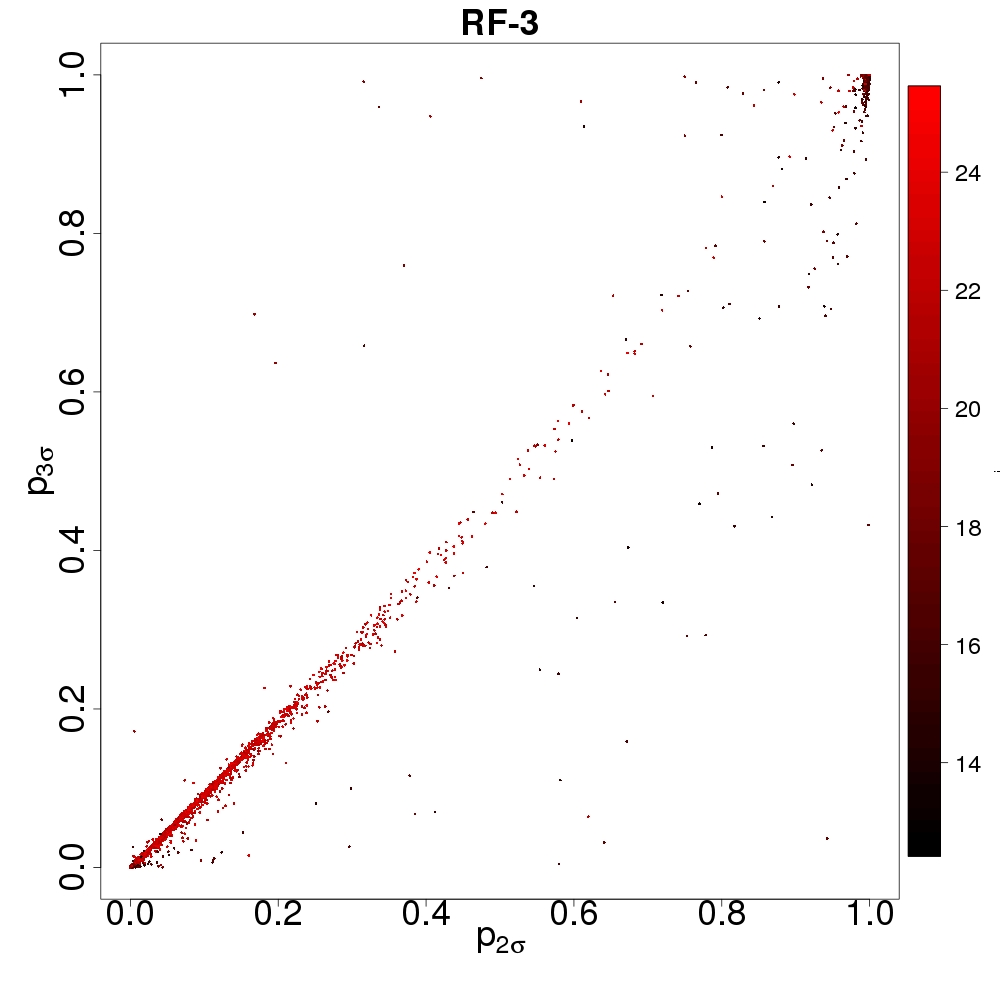}}
  \caption{\label{psigmas}Scatter plots of the membership
    probabilities for the AM (left), RF-2 (middle), and RF-3 (right)
  feature sets, and for two values of the admission probability
  threshold $p_{in}$, namely 0.9545 (2$\sigma$ detection, $x$-axis)
  and 0.9975 (3$\sigma$ detection, $y$-axis). The colour code
    reflects the apparent i-band magnitude according to the colour
    scale to the right.}
\end{figure*}

Figure \ref{psigmas} shows the membership probabilities in the three
representation spaces (AM, RF-2, and RF-3, from left to right)
obtained with two admission thresholds corresponding to 2- and
3-$\sigma$ detections ($x$- and $y$- axes, respectively). The colour
code reflects the apparent i-band magnitude on a linear scale and
shows no systematic effect for points away from the diagonal. We see
that the AM scatter plot portrays a significant lack of robustness,
with a large proportion of sources changing the membership probability
significantly. The RF-2 and RF-3 representation spaces, in contrast,
show a reasonable stability in the membership probabilities assigned,
with the RF-2 showing a more stable behaviour in the region of
$p$=1.

\begin{figure*}[thb]
  \centering\subfigure[]{\label{rf23i}
    \includegraphics[scale=0.17]{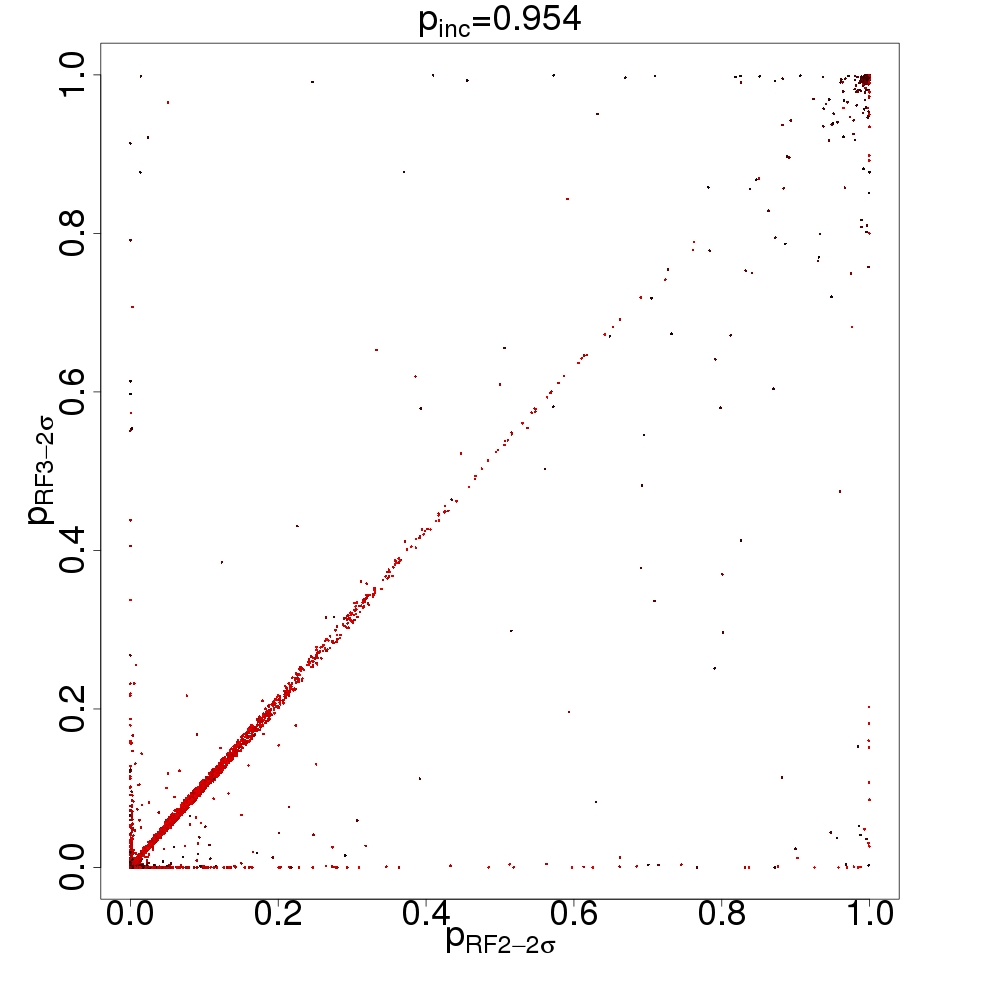}}
  \centering \subfigure[]{\label{rf23ii}
    \includegraphics[scale=0.17]{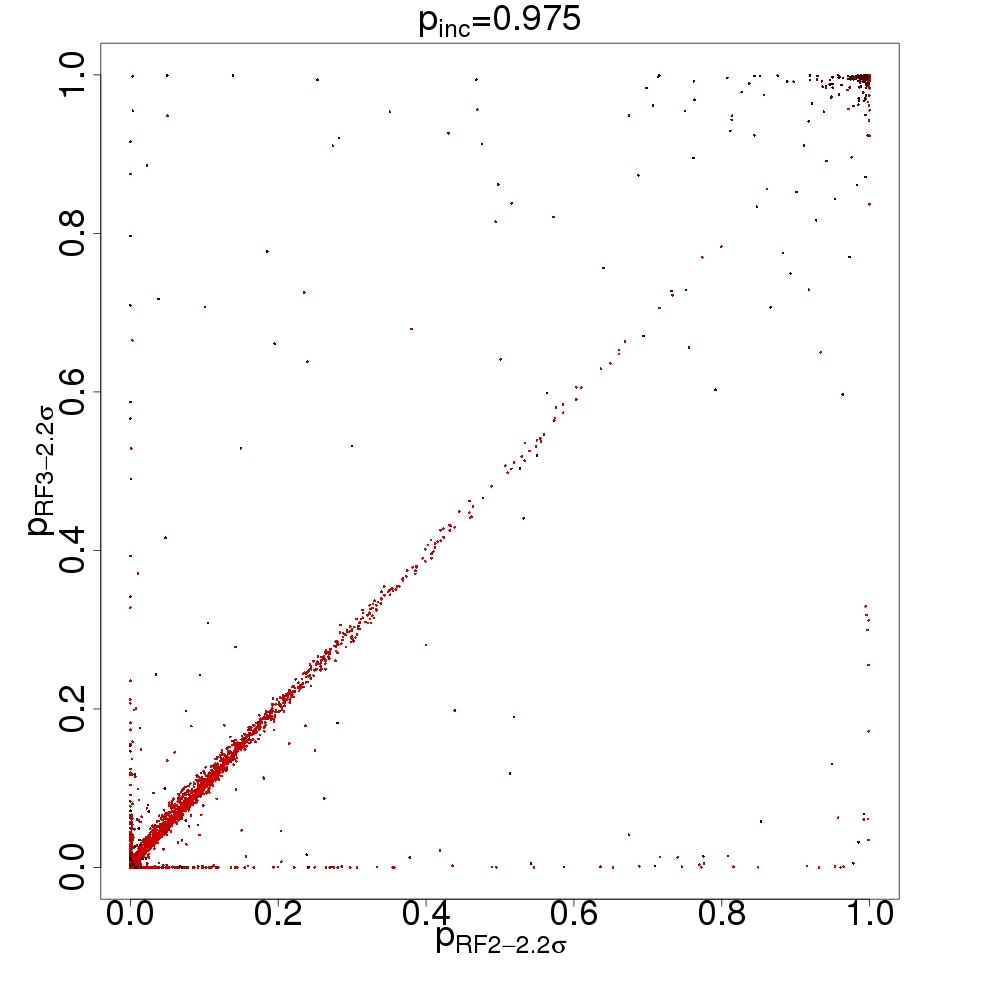}}
  \centering\subfigure[]{\label{rf23iii}
    \includegraphics[scale=0.17]{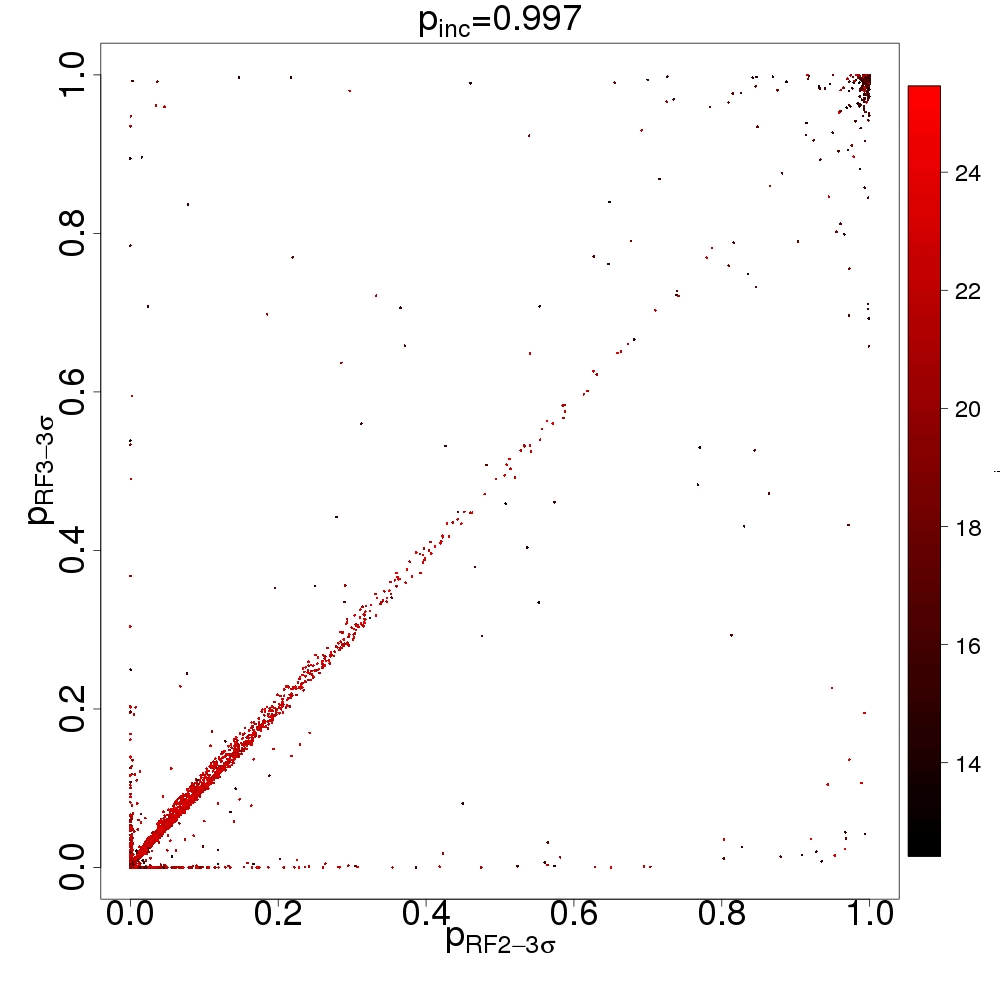}}
  \caption{\label{rf23}Scatter plots of the membership probabilities
    for the RF-2 versus RF-3 representation spaces for three values of
    the admission probability threshold $p_{in}=$ 0.9545 (2$\sigma$
    detection, left), 0.975 (2.2$\sigma$ detection, middle), and 0.9975
    (3$\sigma$ detection, right). The colour code reflects the
    apparent i-band magnitude according to the colour scale to the
    right.}
\end{figure*}

Figure \ref{rf23} shows the correlation between the membership
probabilities obtained in the RF-2 and RF-3 representation spaces for
the three values of $p_{in}$ tested in the experiments.  The colour
scale represents the i-band magnitude on a linear scale from black
(i$\approx$13) to red (i $\approx$ 25) and shows no systematic trend
for the deviations from the diagonal. We find evidence of membership
probabilities obtained in the two representation spaces that strongly
disagree. This is to be expected because a source can be in a region
of high probability density in the variables of one representation
space, but be an outlier in a subspace of the other, especially if the
source has not been observed in a large part of the common feature
space (that is, in the subspace of features common to RF-2 and RF-3),
or if some photometric measurements are problematic (e.g. affected by
cosmic rays or bad pixels) in the set of magnitudes specific to a
feature space. But these are not a large or significant portion of the
total.

From Figs. \ref{cdf-pc-all}, \ref{psigmas}, and \ref{rf23} it is
evident that the various approaches convey slightly different
membership lists. This is an problem that is usually not dealt with in
the literature: the fact that any membership analysis depends (amongst
other factors) on the representation space used and on the membership
threshold adopted. Because we are aware that from a scientific
perspective, different objectives require different samples, we
provide a series of catalogues, one for each experiment
described. This way, the reader can decide which experimental setup
best suits his/her needs. It must be kept in mind that any particular
choice will always be a compromise between contamination (which may
increase with decreasing membership probability thresholds) and
completeness (reduced if very strict criteria are adopted). However,
the relative stability of the RF-2 membership probabilities with
increasing admission threshold, and the better coverage of faint
sources visible in Fig \ref{sensitivity-fig} two desirable properties
that make it a better choice than the other two feature sets. Neither
the reference set defining the likelihood model, nor the distribution
of final membership probabilities obtained with this feature set
change much with the admission threshold $p_{in}$, and thus, we do not
recommend a particular set of probabilities. Figure \ref{rf23} shows
that the high-probability sources are the same in the two
representation spaces, except for a few sources. These discrepancies
can be explained by the fact that RF-3 includes one magnitude that is
not present in RF-2: the Z-band magnitude. This implies that a source
with a high membership probability in the RF-2 representation space,
but with a Z-band magnitude that places it at a large Mahalanobis
distance from the principal curve, can have a significantly lower
membership probability in the RF-3 space.

Finally, we briefly analyse the sensitivity of the membership lists
with respect to the {\em ad hoc} choice of proportions (80/20) for the
parallel sequence of equal-mass binaries. We checked post-hoc the
proportion of sources with higher probability to belong to this
parallel sequence. We did this for complete sources in the final
definition set of the RF-2 representation space and a $p_{in}$
threshold equal to 0.975. Given the final principal-curve model for
the likelihood in the CM subspace and its parallel replicate 0.75
magnitudes above, we computed for each source in the definition set
the probability that each of the two parallel sequences generated the
source, taking into account the measurement uncertainties. We assumed
that the source was generated by the sequence that produces the higher
probability. The resulting proportion of 82.2/17.8 agrees reasonably
well with the input priors. The methodology is fairly robust because
the proper motion measurements strongly constrain the CM subspace
model. As a final check of the robustness of our results, we ran
exactly the same model (in the RF-2 representation space and with the
same $p_{in}$ probability threshold for admission of members to the
definition set) with a 50/50 prior probability for the two
sequences. The resulting definition set was qualitatively and
quantitatively identical except for one source in the 80/20 set that
is missing from the 50/50 set. As a consequence, the principal-curve
likelihood model is indistinguishable in the two cases, and so are the
resulting membership probabilities for the incomplete sources,
rendering the results virtually independent (to within reasonable
limits) of the choice of this prior proportion.

\section{\label{method} Method limitations and conclusions}
\label{sect:lim-concl}

We extended the classical procedure to infer membership probabilities
to deal with proper motions and multiband photometry on an equal basis
and consistently across the full magnitude range. We constructed a
maximum-likelihood, unique multidimensional probabilistic model using
the EM algorithm and an initial reference set of sources assumed to be
cluster members.

There are, however, a few caveats and limitations that must be made
clear to understand how the membership probabilities can be used in
future studies.

In the first place, we did not include incomplete sources in the
derivation of the likelihood models. This is a valuable source of
information that can be incorporated in a consistent derivation of the
model parameters from the complete data set.

We also incorporated the measurement uncertainties in the inference of
the source membership probability (as described in
Eq. \ref{uncertainties}), but we did not include them in the
derivation of the probabilistic models. In fact, the covariance matrix
used for any given value of $\lambda$ (the parameter along the
principal curve), was calculated from the set of sources that project
to that value of $\lambda$, and thus is an overestimate of the true
isochrone width, as explained in Sect. \ref{PC}. Had we included the
measurement uncertainties in the derivation of the likelihood model,
the resulting covariance matrices in $p(m_{CM}|\lambda)$ would
have been narrower and the membership probability estimates lower for
sources far from the isochrone.

Furthermore, we did not infer the binarity fraction in the
principal-curves model for the photometric subspace. We fixed values
{\sl ab initio} that were assumed to be constant all along the
sequence, which does not apply in reality, as shown in Figs.,
\ref{plot-st3} and \ref{plot-missings-st3}. Nor did we explicitly
model the contribution of unequal mass binaries to the probability
density of sources in the representation space (although these systems
are recovered as members due to the overestimation of the covariance
of the models).

Finally, we have no estimate of the uncertainties in the inferred
membership probabilities themselves, a direct consequence of using a
maximum-likelihood model. Neither have we uncertainties on derived
distributions such as the density of sources along the principal curve
(a proxy to the IMF). We are currently working on the construction of
a hierarchical model \citep[see e.g.][and references therein for
  applications in the field of
  astrophysics]{gelman2007data,2010ApJ...725.2166H} from the DANCe
data set for the Pleiades and other young stellar clusters. In this
framework of multilevel or hierarchical models, the observations
(proper motions and multiband photometry in our case, but this can be
generalised to any kind of data) are treated as outcomes drawn from
probabilistic distributions, the parameters of which are, in turn,
drawn from higher-level probability distributions (which include the
IMF, or the star formation rate and the IMF if stellar evolution
models are assumed). The parameters of these higher-level
distributions are known as hyper-parameters. In our case (and just as
an illustrative example of the power of these models), instead of
using a maximum-likelihood estimate of the probability density of
sources along the principal curve, we parameterise the density along
the curve, and include these new parameters into the set to be
inferred from the data. This way, we can derive a
posterior-probability distribution for the biased and censored
luminosity function (LF) and obtain uncertainty estimates or
confidence intervals for its value at a given position along the
sequence. This can be achieved by propagating the measurement
uncertainties in a consistent way, upwards along the various levels of
the hierarchy. Of course, to take full advantage of the power of
hierarchical models, we can go beyond the observed, biased and
censored LF and try to infer the unbiased, uncensored LF up to the
highest observed luminosity (beyond that point, and due to the lack of
observational constraints, the uncertainties on the model parameters
increase). This, however, comes at the cost of modelling the
observational biases, and potentially complex and poorly known
distributions such as that of the reddening or the 3D spatial density
of sources.

In future articles of this series we will analyse the consequences of
the membership probabilities derived here on the distributions of
properties of the Pleiades sources and compare these distributions
with previous studies of the cluster.


\begin{acknowledgements}

H. Bouy is funded by the the Ram\'on y Cajal fellowship program number
RYC-2009-04497. H. Bouy acknowledges funding and support of the
Universit\'e Joseph Fourier 1, Grenoble, France. This research has
been funded by Spanish grants AYA2012-38897-C02-01,
AyA2011-24052, AYA2010-21161-C02-02, CDS2006-00070 and
PRICIT-S2009/ESP-1496. E. Moraux ackowledges funding from the Agence
Nationale pour la Recherche program ANR 2010 JCJC 0501 1 ``DESC
(Dynamical Evolution of Stellar Clusters)''. J. Bouvier acknowledges
funding form the Agence Nationale pour la Recherche program ANR 2011
Blanc SIMI 5-6 020 01 (``Toupies''). J. Bouvier and E. Moraux
acknowledge support from the Faculty of the European Space Astronomy
Centre (ESAC). E. Bertin acknowledges partial funding of computer
resources by the French Programme National de Cosmologie et Galaxies
and CNRS-Fermilab contract \#367561. This work has made an extensive
use of Topcat
\citep[\url{http://www.star.bristol.ac.uk/~mbt/topcat/},][]{2005ASPC..347...29T}.

\end{acknowledgements}

\bibliographystyle{aa} \bibliography{sarro}

\end{document}